\documentclass[12pt,english]{article}
\usepackage{babel}
    \title{Number Operator Algebras}
    \author{Fabien Besnard}

\def\cc#1{\hfill\kern .7em#1\kern .7em\hfill}
\def\r{{\rightarrow}}
\def\l{{\leftarrow}}
\def\End{{\rm End}}
\def\CC{{\rm\bf C}}
\def\RR{{\rm\bf R}}
\def\ZZ{{\rm\bf Z}}
\def\NN{{\rm\bf N}}

\def\II{{\cal I}}
\def\JJ{{\cal J}}
\def\KK{{\cal K}}

\def\MM{{\cal M}}
\def\SS{{\cal S}}
\def\Cl{{\rm Cl}}
\def\aix{{a^+_i}}
\def\ai{{a^{}_i}}
\def\ajx{{a^+_j}}
\def\aj{{a^{}_j}}
\def\akx{{a^+_k}}
\def\ak{{a^{}_k}}

\def\a1{{a^{}_1}}
\def\ax1{{a^+_1}}
\def\az{{a^{}_0}}
\def\azx{{a^+_0}}
\def\aiz{{a_{i_0}^{}}}
\def\aizx{{a_{i_0}^+}}
\def\ajz{{a_{j_0}^{}}}
\def\ajzx{{a_{j_0}^+}}
\def\A2{{a^{}_2}}
\def\Ax2{{a^+_2}}
\def\an{{a^{}_n}}
\def\anx{{a^+_n}}

\def\b1{{b^{}_1}}
\def\bx1{{b^+_1}}
\def\bix{{b^+_i}}
\def\bjx{{b^+_j}}
\def\bi{{b^{}_i}}
\def\bj{{b^{}_j}}
\def\b1{{b^{}_1}}
\def\bx1{{b^+_1}}

\def\aii{{a^{}_\II}}
\def\ajj{{a^{}_\JJ}}
\def\akk{{a^{}_\KK}}

\def\aiix{{a^+_\II}}
\def\ajjx{{a^+_\JJ}}

\def\ssi{\Leftrightarrow}
\def\dem{{\bf Proof : }}
\def\limind{\mathop{\oalign{{\rm lim}\cr\hidewidth$\longrightarrow$\hidewidth\cr}}\limits}
\def\limproj{\mathop{\oalign{{\rm lim}\cr\hidewidth$\longleftarrow$\hidewidth\cr}}\limits}
\def\limtens{\limind_{\JJ\subset\II}{\mathop{\bigotimes\kern
        0pt}_{\JJ}}}
\def\hattens{\mathop{\hat{\bigotimes\kern 0pt}}\limits}
\def\tens{\mathop{\bigotimes\kern 0pt}\limits}
\def\limtens{\limind_{\JJ\subset\II}{\tens_{\JJ}}}

\def\hfl#1#2{\smash{\mathop{\hbox to
      12mm{\rightarrowfill}}\limits^{\scriptstyle#1}_{\scriptstyle#2}}}
\def\vfl#1#2{\llap{$\scriptstyle #1$}\left\downarrow\vbox to
    6mm{}\right.\rlap{$\scriptstyle #2$}}
\def\diagram#1{\def\normalbaselines{\baselineskip=0pt\lineskip=10pt\lineskiplimit=1pt}\matrix{#1}}

\def\be{\begin{equation}}
\def\ee{\end{equation}}
\def\bea{\begin{eqnarray}}
\def\eea{\end{eqnarray}}
\newtheorem{theorem}{Theorem}
\newtheorem{lemma}{Lemma}
\newtheorem{corollary}{Corollary}
\newtheorem{definition}{Definition}
\newtheorem{proposition}{Proposition}

\begin{document}
%\hsize=16.0truecm\vsize=22.5truecm\vglue6.3truecm
\vspace{5cm}
\maketitle
\begin{abstract}
Under some hypotheses (symmetry, confluence), we enumerate all quadratically presented algebras, generated by creation and destruction operators, in which number operators exist. We show that these are algebras of bosons, fermions, their immediate generalizations that we call pseudo-bosons and pseudo-fermions, and also matrix algebras, in the finitely generated case. We then recover $q$-bosons (and pseudo-$q$-bosons) by a completion operation.

\end{abstract}
\eject

\section{Introduction}
In \cite{bes1} we have proposed a new way of looking at quantization. In this point of view, one should quantize the equations of evolution rather than the canonical commutation relations, the latter being a consequence of the former. In the case of a system of harmonic oscillators, which is crucial for field theory, we want to find algebras (over $\CC$) of q-numbers generated by a set $\{\ai|i\in\II\}$ of so-called destruction operators, and a set $\{\aix|i\in\II\}$ of creation operators, conjugate to the former by an anti-involution which we denote by $J$. Our algebras will then have the structure of $*$-algebras. It should be stressed that the word ``operator'' is just a convention here, since no Hilbert space is a priori fixed. We require the existence of elements $N_i$, for all $i\in\II$, so that the following equations hold :
\be
[N_i,\aj]=-\delta_{ij}\ai\label{eq1}
\ee
\be
[N_i,\ajx]=\delta_{ij}\aix\label{eq2}
\ee

Since the base field $\CC$ does not play a particular role, we will replace it by any field of characteristic 0.

In this article we propose to give a detailed account of the results obtained in our thesis, as well as a few novelties concerning Fock algebras. We will begin in the second section by defining precisely what we call a number operator algebra, and work out the first consequences of the definition. In the third section we will restrict to the case of quadratically confluent number operator algebras of finite type, and state the classification theorem. The proof is quite long, so we will not give it here in full. However we give an account of the demonstration, as detailed as we can, in section four. Then, we will tackle to the problem of n.o.a. of infinite type in section five. In particular we will show that we can do without the confluence hypothesis. Furthermore, only four out of the six different kinds of algebras we have found in the finite case remain. These algebras are precisely those which can be obtained as deformations of $\epsilon$-Poisson algebras (a generalization of Super-Poisson algebras about which one can consult \cite{sch} or \cite{bes2}). In section six, we will prove a generalization of the classification theorem for n.o.a. of infinite type, in which we let the number operators belong to a certain completion of the algebra. We will see that this completion operation is natural in order to have a Fock representation. In this case, we will find two more solutions, namely $q$-bosonic and pseudo-$q$-bosonic algebras.

\section{Definitions and first consequences}
\subsection{Definitions}
We fix once for all $K$ a field of characteristic 0, $\tau$ an involution of $K$ (possibly the identity), and $R$ the sub-field of elements of $K$ fixed by $\tau$. For each cardinal number $\alpha$ we choose a representative set $\II_\alpha$. In particular if $\alpha=n$ is finite, we take $\II_n=\{1,\ldots,n\}\subset\NN$. All algebras are unital $K$-algebras and all morphisms preserve units. If $A$ is an algebra, we say that it is trivial iff $A=0$ or $A=K$. We denote by $Z(A)$ the center of $A$.
\begin{definition}
Let $\alpha$ be a cardinal number and $B$ be a non-trivial $K$-algebra. Let $X_A=\{\ai|i\in\II_\alpha\}$, $X_{A^+}=\{\aix|i\in\II_\alpha\}$, $N=\{N_i|i\in\II_\alpha\}$ be 3 sets indexed by $\II_\alpha$, with the $\ai$'s and $\aix$'s in $B$, and $N_i$'s in $B/Z(B)$. We call $(B,X_A,X_{A^+},N)$ a number operator algebra of type $\alpha$ if, and only if :

\item{(i)} $B$ is generated by $X_A\cup X_{A^+}$ as an algebra.
\item{(ii)} One uniquely defines an anti-involution $J$ on $B$ by setting $J(\ai)=\aix$.
\item{(iii)} Equations (\ref{eq1}) and (\ref{eq2}) are fulfilled.
\end{definition}

\underline{Remark} : $N_i$ belongs to $B/Z(B)$ which is only a vector space, nevertheless the commutator of such an element with any element of $B$ is well defined, so that equations (\ref{eq1}) and (\ref{eq2}) make sense.

We have to define morphisms between two n.o.a. : we will only need to do so for n.o.a. of the same type. For a more general definition, see \cite{bes1}.
\begin{definition}
Let $(B,X_A,X_{A^+},N)$ and $(B',X_{A}',X_{A^+}',N')$ be two n.o.a. of type $\alpha$, and let $f$ be an algebra homomorphism from $B$ to $B'$. We will say that $f$ is a morphism of n.o.a. iff :
	\item{(i)} $f\circ J=J'\circ f$
	\item{(ii)} $f(Z(B))\subset Z(B')$
	\item{(iii)} There exists a bijection $\phi :\II_\alpha\rightarrow\II_\alpha$ such that $f(N_i)\in N_{\phi(i)}'+Z(B')$
\end{definition}

It is easy to see that n.o.a. of type $\alpha$ and their morphisms form a category.

We are now going to define another category, in which all creation (resp. destruction) operators play a symmetric role. We will denote by $\SS_\alpha$ the group of permutations of $\II_\alpha$ that leave all but a finite number of elements invariant.

\begin{definition}
Let $B$ be a n.o.a. of type $\alpha$ and $\SS_\alpha$ be the group of permutations of $\II_\alpha$ with finite support. We will say that $B$ is symmetric iff for all $\sigma\in\SS_\alpha$ the prescription $\ai\mapsto a_{\sigma(i)}$ uniquely defines an automorphism $\sigma^*$ of $B$ commuting with $J$.
\end{definition}

\begin{definition}
A n.o.a. morphism $f$ between two symmetric n.o.a. of type $\alpha$ will be called symmetric iff it commutes with $\sigma^*$, $\forall\sigma\in\SS_\alpha$.
\end{definition}
Let $(B,X_A,X_{A^+},N)$ be a n.o.a. of type $\alpha$, set $X=X_A\cup X_{A^+}$, and let $\langle X\rangle$ be the free monoid, $L_\alpha=K\langle X\rangle$ the free algebra, generated by $X$. Thus, the elements of $L_\alpha$ are of the form $x=\sum_i\lambda_i x_i$, with $\lambda_i\in K$, $x_i\in\langle X\rangle$. The set of those monomials $x_i$ such that $\lambda_i\not=0$ is called the support of $x$, it is a finite set. The $\lambda_ix_i$'s are called the terms of $x$.

We call $\pi$ the canonical projection from $L_\alpha$ onto $B$, and $I$ the kernel of $\pi$.

There is an anti-involution on $L_\alpha$ sending $\ai$ to $\aix$. We also denote it by $J$. The symmetric group $\SS_\alpha$ also acts on $L_\alpha$ in an obvious way. If $B$ is symmetric this action commutes with $\pi$. Since everything is commuting with $\pi$ we will often drop it from the notations : whether an element belongs to $L_\alpha$ or $B$ should be clear from the context. 

\begin{definition}
If $I$ is generated by quadratic elements we say that $B$ is a quadratic n.o.a.
\end{definition}

\underline{Remark} : What we call a quadratic element is an element of degree $\leq 2$. If it has no term of degree $\leq 1$ we call it homogenous quadratic.

\subsection{A few lemmas}
We denote by $\ZZ^{(\alpha)}$ the direct sum $\bigoplus_{i\in\II_\alpha}\ZZ$. Its elements are mappings $p : \II_\alpha\rightarrow\ZZ$ with a finite support.

For all $i\in\II_\alpha$ we define the derivation ${\cal N}_i : L_\alpha\rightarrow L_\alpha$ on the generators by ${\cal N}_i(\aj)=-\delta_{ij}\ai$ and ${\cal N}_i(\ajx)=\delta_{ij}\aix$. The following trivial lemma has important consequences :
\begin{lemma}\label{lem1}
If $B$ is a n.o.a. then the diagram of vector spaces : 
$$\diagram{
L_\alpha & \hfl{{\cal N}_i}{} & L_\alpha \cr
\vfl{\pi}{} & & \vfl{\pi}{}\cr
B & \hfl{ad(N_i)}{} & B \cr
}$$
commutes for all $i\in\II_\alpha$.
\end{lemma}
\begin{corollary}\label{cor1}
$B$ is a $\ZZ^{(\alpha)}$-graded algebra. More precisely : ${\displaystyle L_\alpha=\bigoplus_{p\in\ZZ^{(\alpha)}}L_\alpha^p}$, ${\displaystyle B=\bigoplus_{p\in\ZZ^{(\alpha)}}B^p}$, ${\displaystyle I=\bigoplus_{p\in\ZZ^{(\alpha)}}(I\cap L_\alpha^p)}$, with $L_\alpha^p=\{x\in L_\alpha|\forall
i\, {\cal N}_i(x)=p(i)x\}$, and $B^p=\{y\in B|ad(N_i)(y)=p(i)y\}$.
\end{corollary}
\begin{corollary}\label{cor2}
$$\forall i,j\quad [N_i,N_j]=0$$
\end{corollary}
\dem\newline
To prove the lemma, one just has to verify that $\pi({\cal N}_i(x))=[N_i,\pi(x)]$ for $x\in X$ since ad$(N_i)=[N_i,.]$ is a (inner) derivation.

For the corollary \ref{cor1} just notice that ${\cal N}_i{\cal N}_j={\cal N}_j{\cal N}_i$ for all $i,j$. Then according to lemma \ref{lem1} we also have $[$ad$(N_i),$ad$(N_j)]=0$, where $[,]$ is the commutator in End$(B)$. Thus we can decompose each space into common eigenspaces for the appropriate family of commuting endomorphisms. 

Let us prove corollary \ref{cor2} : $\forall x\in B$,
$0=[$ad$(N_i),$ad$(N_j)](x)=$ad$([N_i,N_j])(x)$. Thus $[N_i,N_j]\in
Z(B)$. But the central elements commute with all $N_i$, thus $Z(B)\subset B^0$. Now, if
$N_j=\sum_{p\in\ZZ^\alpha}N_j^p$, then $\forall i$ :
$$[N_i,N_j]=\sum_{p\in\ZZ^\alpha}p(i)N_j^p\in B^0$$
$$\Rightarrow \forall i,\forall p\not=0,\quad p(i)N_j^p=0$$
%$$\Rightarrow \forall p\not=0,\quad N_j^p=0$$
$$\Rightarrow N_j\in B^0\Rightarrow [N_i,N_j]=0$$
\hfill QED.

Let us notice that the action of the derivations ${\cal N}_i$ (resp. $[N_i,.]$) on a monomial $x$ is to multiply it by the integer $n_i(x)$, which is the number of $\aix$ minus the number of $\ai$ appearing in $x$. We will call $n_i(x)$ the {\it i-number} of $x$. More generally if $x$ belongs to an eigenspace of ${\cal N}_i$ or ad$(N_i)$ we call the corresponding eigenvalue the $i$-number of $x$.

\begin{lemma}
If $B$ is symmetric, $B^0$ is a sub-representation space for $\SS_\alpha$.
\end{lemma}
\dem\newline
Indeed, if $x\in B^0$ it is a linear combination of monomials of zero $i$-number for all $i$, and for $\sigma\in\SS_\alpha$, $\sigma(x)$ shares the same property.\hfill QED.

\section{Number Operator Algebras of Finite Type}
\subsection{The confluence hypothesis}
In order to state the fundamental confluence hypothesis, we have to introduce some combinatorial terminology. In this subsection we do not have to suppose yet that $\alpha$ is finite, although we will only need the confluence hypothesis in this case. We refer the reader to \cite{Berg} or \cite{ufn} for a more formal presentation. In the sequel by a {\it reduction system} we mean a subset of $\langle X\rangle\times L_\alpha$. Its elements are called reductions : they are couples $(m,f)$ for which we will use the notation $m\rightarrow f$.

Given a presentation of an ideal $I$ (i.e. a set of generators) and a monoid ordering $<$, that is an ordering on $\langle X\rangle$ which is compatible with multiplication, it is sometimes possible (always if $<$ is total) to construct a reduction system by isolating the leading monomial of every element of the presentation. We say that this reduction system is associated with the presentation and $<$. More precisely, if $P$ is the presentation, then the reduction system associated with $P$ and $<$ is $S_{I,<}=\{$lm$(g)\rightarrow -{1\over{\rm lc}(g)}(g-$lt$(g))|g\in P\}$ where we used the following notations : lm stands for ``leading monomial'', lc stands for ``leading coefficient'' and lt stands for ``leading term''.

A reduction system is useful if it gives a way of rewritting elements of $B$ so as to give them a unique normal form. Indeed, let $x=\sum_i\lambda x_i$ be an element of $L_\alpha$. If $m\rightarrow f$ is a reduction of $S_{I,<}$, then every occurence of $m$ as a subword of any monomial $x_i$ may be replaced by $f$ without changing the class of $x$ modulo $I$. The aim is then to apply every possible reduction to $x$ until we get an {\it irreducible} element, that is to say an element we cannot reduce any further.

Of course this is not always a well defined procedure. First if we have two reductions $m\rightarrow f$ and $m'\rightarrow f'$ it can happen that the same monomial $x_i$ can be written $x_i=abc$ with $ab=m$ and $bc=m'$. In this case we say that there is an {\it overlap ambiguity}. It is called solvable if there are two sequences of reductions, $s_1$ and $s_2$, such that applying $s_1$ on $fc$ and $s_2$ on $af'$ gives the same result. This can be visualized on the following diagram :
$$\matrix{
 &abc&  \cr
\swarrow& &\searrow\cr
fc\hfill& &\hfill af'\cr
{\scriptstyle s_1}\searrow&&\swarrow {\scriptstyle s_2}\cr
 &\mbox{same result}&\cr
}$$
There can also be {\it inclusion ambiguities} : $x_i=abc$ with $abc=m$ and $b=m'$. It is said to be solvable if there are two sequences of reductions $s_1$ and $s_2$ such that $s_1$ applied on $f$ is equal to $s_2$ applied on $af'b$.

When all ambiguities are solvable, the reduction system is said to be {\it confluent}.

There is one last problem to solve : we must be sure that the procedure will stop, and will not give an infinite cycle of reductions. This is achieved by using orderings satisfying the descending chain condition (DCC) : all decreasing sequences are stationnary. Among such orderings, the most natural ones are the so-called ``deglex'' (degree-lexicographic) orderings, obtained from a total ordering $<_0$ on the generators, that is : $x<y$ iff $d^\circ(x)<d^\circ(y)$ or ($d^\circ(x)=d^\circ(y)$ and $x$ is before $y$ in the lexicographic order induced by $<_0$).

So, if $S_{I,<}$ is confluent and if $<$ is a monoid ordering satisfying DCC, Bergman's diamond lemma \cite{Berg} states that the set of irreducible monomials is a $K$-basis for $B$.

For instance take $\alpha=1$, $X_A=\{a\}$, $X_{A^+}=\{a^+\}$, so that $L_1=K\langle a,a^+\rangle$, and denote by $<$ the deglex-ordering coming from $a^+<_0a$. Let us consider the ideal $I$ generated by $P=\{a^2,{a^+}^2,aa^++a^+a-1\}$. The reduction system associated to $P$ and $<$ is $S=\{a^2\rightarrow 0,{a^+}^2\rightarrow 0,aa^+\rightarrow 1-a^+a\}$. This system is easily seen to be confluent. For instance the overlap ambiguity coming from $a^2a^+$ is solvable because $(a^2)a^+\rightarrow 0$ and $a(aa^+)\rightarrow a(1-a^+a)=a-(aa^+)a\rightarrow a-(1-a^+a)a=a^+(a^2)\rightarrow 0$. By Bergman's lemma we find that the irreducible monomials ($1$, $a$, $a^+$ and $a^+a$) form a $K$-basis of $B=L_1/I$.

\underline{Remark :} It is always possible to avoid inclusion ambiguities in a reduction system (see [Berg] or [Bes1]). In this case we say that the reduction system is simplified. It is also always possible to assume that every element of a confluent reduction system is of the form $m\rightarrow r$ with $r$ irreducible. We shall say that such a reduction system is reduced.

\begin{definition}
We say that a presentation $P$ of an ideal $I$ is quadratically confluent (resp. deglex-quadratically confluent) iff the elements of $P$ are at most of degree two, and there exists a monoid ordering $<$ satisfying DCC (resp. a deglex ordering), such that the reduction system associated with $P$ and $<$ is confluent. 
\end{definition}

\subsection{The Main Theorem}
We can now state our main result for n.o.a. of finite type :

\begin{theorem}\label{theo1}
Let $n$ be a finite number and let $B=L_n/I$ be a symmetric deglex-quadratically confluent n.o.a., i.e. $I$ satisfies the following properties 
\item{$(P_0)$} $I\not=L_n$, $I\not=\langle X\rangle$.
\item{$(P_1)$} $J(I)\subset I$.
\item{$(P_2)$} $\forall\sigma\in{\cal S}_n$, $\sigma^*(I)\subset I$.
\item{$(P_3)$} $\exists N_1,\ldots,N_n\in B$ s.t. (\ref{eq1}) and (\ref{eq2}) hold.
\item{$(P_4)$} : $\exists <_0$, a total ordering on $X$ s.t. $I$ admits a quadratic and confluent reduction system, adapted to the deglex ordering coming from $<_0$.
\medbreak
then, if $n=1$, there exists  $h\in R\setminus\{0\}$ such that $I$ is generated by one of the following sets :
\item{(a)} $\{a^2,{a^+}^2,aa^++a^+a-h\}$
\item{(b)} $\{aa^+-a^+a-h\}$

if $n\geq 2$ there exists $h\in R\setminus\{0\}$ such that $I$ is generated by one of the following sets :
\item{(a)} $\{\ai^2,\aix^2,\ai\aj+\aj\ai,\aix\ajx+\ajx\aix,\ai\ajx+\ajx\ai,\ai\aix+\aix\ai-h| 1\leq i\not=j\leq n\}$
\item{(a')} $\{\ai^2,\aix^2,\ai\aj-\aj\ai,\aix\ajx-\ajx\aix,\ai\ajx-\ajx\ai,\ai\aix+\aix\ai-h|1\leq i\not=j\leq n\}$
\item{(b)} $\{\ai^2,\aix^2,\ai\aj,\aix\ajx,\ai\ajx,\ai\aix+\sum_k\akx\ak-h|1\leq i\not=j\leq n\}$
\item{(b')} $\{\ai^2,\aix^2,\ai\aj,\aix\ajx,\aix\aj,\aix\ai+\sum_k\ak\akx-h|1\leq i\not=j\leq n\}$
\item{(c)} $\{\ai\aj-\aj\ai,\aix\ajx-\ajx\aix,\ai\ajx-\ajx\ai,\ai\aix-\aix\ai-h|1\leq i\not=j\leq n\}$
\item{(c')} $\{\ai\aj+\aj\ai,\aix\ajx+\ajx\aix,\ai\ajx+\ajx\ai,\ai\aix-\aix\ai-h|1\leq i\not=j\leq n\}$

\end{theorem}

\underline{Remark 1} : In the case $n=1$, the hypothesis $(P_4)$ can be loosened to :
\smallbreak
{\hfill $I$ is generated by elements of degree $\leq 2$\qquad$(P_4')$\hfill}
\smallbreak
It is also true for $n$ infinite that we can replace $(P_4)$ with $(P_4')$, as we shall see later. However, in the case $2\leq n<\infty$ there exist ideals $I$ satisfying $(P_0),\ldots,(P_3)$ and $(P_4')$ but not $(P_4)$ (see \cite{bes1}).

\underline{Remark 2} : In the physical case, $h$ is a positive real number and we can set $h$ to $1$ by rescaling the units, which amounts to the symmetric n.o.a. isomorphism $\phi_\lambda : \ai\mapsto\lambda\ai$, with $\lambda\in\RR$. Then, we get :
	\begin{itemize}
	\item{(a)} The tensor product of $n$ Weyl algebras (boson case), $A_n=A_1\otimes\ldots\otimes A_1$, with $A_1=L_1/\langle aa^+-a^+a-1\rangle$.
	\item{(a')} The graded tensor product of $n$ Weyl algebras (we call it the pseudo-boson case), $\hat A_n=A_1\hat\otimes\ldots\hat\otimes A_1$.
	\item{(b)} The matrix algebra $\MM_{n+1}(K)$.
	\item{(b')} The same as above but with the creation and destruction operators exchanged.
	\item{(c)} The graded tensor product of $n$ Clifford algebras (the fermion case), $\hat C_n=C_1\hat\otimes\ldots\hat\otimes C_1$, with $C_1=L_1/\langle a^2,{a^+}^2,aa^++a^+a-1\rangle$.
	\item{(c')} The tensor product of $n$ Clifford algebras (the pseudo-fermion case), $C_n=C_1\otimes\ldots\otimes C_1$. 
	\end{itemize}
Particles whose creation and destruction operators form the algebra $(c)$ or $(c')$ satisfy Pauli's exclusion principle : only one particle of that kind can be found in a given state ($\ai^2=\aix^2=0$). Particles of type $(b)$ or $(b')$ follow a more extreme exclusion principle : only one such particle can be found, regardless of its state ($\ai\aj=\aix\ajx=0$).

\underline{Remark 3} : We see that all these algebras depend only on a single constant $h$. Thanks to this fact we can see them as deformations of the ``classical'' algebras obtained by taking $h=0$. This point of view is developped in \cite{bes2}.

\section{Sketch of proof of theorem \ref{theo1}}
We do not have the space here to give the full proof. Nevertheless, we will give enough indications (we hope) for the reader to fill in the blanks. For a detailed proof, see \cite{bes1}.
\subsection{A few more lemmas} 
\begin{lemma}\label{lem2}
If $(P_3)$ and $(P_4')$ are satisfied, then $I$ can be generated by a set of elements of the form : $\ai^2$ (1), $\aix^2$ (1'), $\alpha\ai\aj+\beta\aj\ai$ (2), $\alpha\aix\ajx+\beta\ajx\aix$ (2'), $\alpha\ai\ajx+\beta\ajx\ai$ (3), ${\sum_{1 \leq i\leq n}\alpha_i\ai\aix+\sum_{1 \leq
    i\leq n}\beta_i\aix\ai-\lambda}$ (4), $\ai$ (5), or $\aix$ (5').
\end{lemma}
\dem\newline
Let $P$ be a quadratic presentation of $I$ and let $r\in P$. We can write $r=\sum_{p\in\ZZ^n} r^p$, where $p$, seen as a function of $i$, must be $\pm 2\delta_{ij}$, or $\pm(\delta_{ij}+\delta_{ik})$, or $0$, or $\pm\delta_{ij}$. Now, from lemma \ref{lem1}, $r\in I\ssi r^p\in I$, $\forall p$, and the forms $(1)$ to $(5')$ correspond to the different possibilities for the $p$'s.\hfill QED.

\begin{lemma}\label{lem3}
If $I$ fulfils $(P_0)$, $(P_3)$, and $(P_4')$, then it must contain at least one set of generators of type (4) with $\lambda\not=0$.
\end{lemma}
\dem\newline
Let us suppose that it is not so. Then, by lemma \ref{lem2} and $(P_4')$, $I$ must be generated by elements of the form : $\ai$, $\aix$, or $r$, with $r$ homogenous quadratic. Now by $(P_0)$, $\exists i$ such that $\ai$ or $\aix$ is not in $I$. Suppose $\ai\notin I$, and let ${\tilde N}_i$ belong to $\pi^{-1}(N_i)$. By $(P_3)$, $I$ must contain ${\tilde N}_i\ai-\ai{\tilde N}_i+\ai$, whose only term of degree one is $\ai$. But every element of $I$ can be written as $\sum xry+\sum s_j\aj t_j+\sum u_k\akx v_k$,
$r$ being homogenous quadratic, and the second sum running over $j\not=i$. This is a contradiction.\hfill QED.
\medbreak
\begin{lemma}\label{lem4}
If $B=L_n/I$ satisfies $(P_0)$, $(P_3)$ and $(P_4')$, and if $C$ is a commutative algebra, then $\mbox{Hom}_{K-alg}(B,C)=\{0\}$.
\end{lemma}
\dem\newline
We have $\forall i$, $\phi([N_i,\pi(\aix)])=\phi(\pi(\aix))=0$, since $C$ is commutative. For the same reason, $\phi(\pi(\ai))=0$. Now, by lemma \ref{lem3}, $I$ contains an element of the form $x+\lambda.1$, $\lambda\not=0$, $x$ homogenous quadratic. We have :
$0=\phi(\pi(x+\lambda.1))=\phi(\pi(x))+\lambda\phi(1)=0+\lambda
.\phi(1)$. Therefore $\phi(1)=0$, and $\phi=0$.\hfill QED.

\begin{lemma}\label{lem5}
Let $B=L_n/I$, $B'=L_n/I'$, $\pi$ and $\pi'$ the respective projections. If $\exists\phi$: $B\rightarrow B'$, an algebra homomorphism such that $\phi(\pi(\ai))=\pi'(\ai)$ and 
$\phi(\pi(\aix))=\pi'(\aix)$, then : $I$ fulfils $(P_3)$ $\Rightarrow$
$I'$ fulfils $(P_3)$.\newline
(In particular, this is the case if $I\subset I'$ and if $\phi$ is induced by the identity map of $L_n$)
\end{lemma}
\dem\newline
It is easily verified that the images by $\phi$ of the number operators of $B$ are number operators for $B'$.\hfill QED.

\begin{lemma}\label{lem6}
Let $B=L_n/I$ such that $(P_3)$ holds, let $B'$ be any algebra, and $\phi\in{\rm Hom}_{K-\rm alg}(B,B')$ such that $\forall i$, $\phi(\ai)=\phi(\aix)$. Then, $\forall i$ $\phi(\ai)=0$.
\end{lemma}
\dem\newline
Set $x_i:=\phi(\ai)=\phi(\aix)$. In one hand $[N_i,\ai]=-\ai\Rightarrow [\phi(N_i),x_i]=-x_i$, and in the other hand $[N_i,\aix]=\aix\Rightarrow [\phi(N_i),x_i]=x_i$. So $x_i=0$.\hfill QED.

\begin{lemma}\label{lem7}
Let $n\geq 2$, $B=L_n/I$ such that $(P_3)$ holds, $B'$ any algebra, and $\phi\in{\rm Hom}_{K-\rm alg}(B,B')$ such that $\forall i,j$, $\phi(\ai)=\phi(\aj)$ (resp. $\phi(\aix)=\phi(\ajx)$). Then, $\forall i$ $\phi(\ai)=0$ (resp. $\phi(\aix)=0$).
\end{lemma}
\dem\newline
Let us examine the first case, the other one being similar. Let $i\not=j$ and let $x:=\phi(\ai)=\phi(\aj)$. Then $[N_i,\ai]=-\ai\Rightarrow [\phi(N_i),x]=-x$, and $[N_i,\aj]=0\Rightarrow [\phi(N_i),x]=0$. Thus $x=0$.\hfill QED.
\medbreak

We now have to work out the consequences of $(P_2)$. It is a bit long, but very easy. We only state the results, leaving the details to the reader (one could also see \cite{bes1}).

In what follows, we suppose $n\geq 2$, $V$ is a $n$ dimensional vector space with basis $e_1,\ldots,e_n$ and coordinates $\epsilon_1,\ldots,\epsilon_n$. Let $\rho : {\cal
  S}_n\rightarrow {\rm End}(V)$ be the representation given by $\rho(\sigma)(e_i):=\sigma.e_i:=e_{\sigma(i)}$, let $H$ be the hyperplane of equation $\epsilon_1+\ldots+\epsilon_n=0$. $W$ is the vector space $W=V\oplus V\oplus{\bf 1}$, where ${\bf 1}$ is the trivial representation of ${\cal S}_n$ of dimension 1. $W$ bears the representation $\rho\oplus\rho\oplus 1$. Finally, we set $x_i=e_i\oplus 0\oplus 0$, $y_i=0\oplus e_i\oplus 0$, and let $1$ be a non-zero vector of ${\bf 1}$.

\begin{lemma}\label{lem9}
Let $w\in W$, $w\not=0$, and let $O(w)$ be the linear span of the orbit of $w$ under the action of $\SS_n$. Then $O(w)$ is isomorphic as a representation space to : ${\bf 1}$, $H$, $H\oplus {\bf 1}$, $H\oplus H$ or $H\oplus H\oplus
  {\bf 1}$. Furthermore $O(w)$ has a basis of the form :
      \begin{itemize}
        \item{in the 1st case} : $\{\mu_x.1_x+\mu_y1_y+\mu_1.1\}$, with
  $1_x=x_1+\ldots+x_n$, $1_y=y_1+\ldots+y_n$, and 
  $\mu_x,\mu_y,\mu_1\in K$, not all zero.
        \item{in the 2nd case} :
  $\{\lambda_x(x_i-x_1)+\lambda_y(y_i-y_1)|i>1\}$,
  $(\lambda_x,\lambda_y)\not=(0,0)$.
        \item{in the 3rd case} :
          $\{\mu_x1_x+\mu_y1_y+\mu_11,\lambda_x(x_i-x_1)+\lambda_y(y_i-y_1)|i>1\}$ $(\mu_x,\mu_y,\mu_1)\not=(0,0,0)$, $(\lambda_x,\lambda_y)\not=(0,0)$.
        \item{in the 4th case} : $\{x_i-x_1,y_i-y_1|i>1\}$.
        \item{in the 5th case} :
          $\{\mu_x1_x+\mu_y1_y+\mu_11,x_i-x_1,y_i-y_1|i>1\}$,
          $(\mu_x,\mu_y,\mu_1)\not=(0,0,0)$.
      \end{itemize}
\end{lemma}
\medbreak

If we take an element $x$ of $I$, we can make $\SS_n$ act upon it to get others, so that the whole orbit of $x$ belongs to $I$, and of course, so does its linear span. Using this and noticing that the last lemma apply to our situation if we set $x_i=\ai\aix$, $y_i=\aix\ai$, and $1=1$, we arrive at the following result :

\begin{lemma}\label{lem10}
If $I$ fulfils $(P_0)$, $(P_1)$, $(P_2)$, $(P_3)$ and $(P_4')$, then $I$ can be generated by a union of sets, each having one of the following forms :
\begin{itemize}
\item{form $(2,0)$ :} $\{\ai^2,{\aix}^2|1\leq i\leq n\}$
\item{form $(1,1)_a$ :}
  $\{\ai\aj+\aj\ai,\aix\ajx+\ajx\aix|i<j\}$
\item{form $(1,1)_b$ :} $\{\ai\aj-\aj\ai,\aix\ajx-\ajx\aix|i<j\}$
\item{form $(1,1)_c$ :} $\{\ai\aj,\aix\ajx|i\not=j\}$
\item{form $(1,-1)_a$ :} $\{r\ai\ajx+s\ajx\ai|i\not=j\}$, with
  $(r,s)\not=(0,0)$, $r,s\in R$.
\item{form $(1,-1)_b$ :} $\{\ai\ajx,\ajx\ai|i\not=j\}$
\item{form $(0,0)$ :}
  $\{{\displaystyle{\sum_i\alpha_i\ai\aix+\sum_i\beta_i\aix\ai-\lambda\}}}$
\end{itemize}
Furthermore, each set of the form (0,0) can be replaced by a union of sets of the form :
\begin{itemize}
\item{form $A_1$ :} $\{\ai\aix-\a1\ax1|i>1\}$
\item{form $B_1$ :} $\{\aix\ai-\ax1\a1|i>1\}$
\item{form $A_2$ :} $\{\ai\aix-\lambda|1\leq i\leq n\}$
\item{form $B_2$ :} $\{\aix\ai-\mu|1\leq i\leq n\}$
\item{form $C$ :} $\{{\displaystyle\sum_{1\leq i\leq
      n}}\ai\aix-\lambda\}$
\item{form $D$ :} $\{{\displaystyle\sum_{1\leq i\leq
      n}}\aix\ai-\mu\}$
\item{form $E_1$ :}
  $\{\alpha(\ai\aix-\a1\ax1)+\beta(\aix\ai-\ax1\a1)|i>1\}$, $\alpha\beta\not=0$
\item{form $F$ :} $\{\alpha{\displaystyle\sum_{1\leq i\leq
    n}}\ai\aix+\beta{\displaystyle\sum_{1\leq i\leq n}}\aix\ai-\lambda\}$, $\alpha\beta\not=0$
\end{itemize}
In every case, we can assume that $\alpha$, $\beta$,
$\lambda$, $\mu$ and $\nu$ belong to $ R$.
\end{lemma}

The next step is to combine the different sets of generators enumerated by lemma \ref{lem10}. For instance, if we are given the set $A_1\cup C$, we can replace it with a set of the form $A_2$. Obviously, some combinations, such as the union of two sets of the form $A_2$ with different values of lambda, give a trivial result and we can get rid of them. The next proposition sum up the different results.

\begin{proposition}\label{prop7}
Let $I_2=\{x\in I|d^\circ(x)\leq 2\}$, $I_2^{(2,0)}=I\cap{\rm Span}\{\ai^2,\aix^2|1\leq i\leq n\}$, $I_2^{(1,1)}=I\cap{\rm Span}\{\ai\aj,\aix\ajx|i\not=j\}$, $I_2^{(1,-1)}=I\cap{\rm Span}\{\ai\ajx,\aix\aj|i\not=j\}$, $I_2^{(0,0)}=I_2\cap L_n^{(0,\ldots,0)}$.\newline
Under the hypotheses of lemma \ref{lem10}, there exists a presentation $R$ for 
$I$, of the form $R=R^{(2,0)}\coprod R^{(1,1)}\coprod R^{(1,-1)}\coprod R^{(0,0)}$, such that $R^{(i,j)}$ is a basis of $I_2^{(i,j)}$. Furthermore :
        \begin{itemize}
                \item{}$R^{(2,0)}=(2,0)$ or the empty set.
                \item{}$R^{(1,1)}=(1,1)_a$ or $(1,1)_b$ or $(1,1)_c$ or $\emptyset$.
                \item{}$R^{(1,-1)}=(1,-1)_a$ or $(1,-1)_b$ or $\emptyset$.
        \end{itemize}
\smallbreak
and $R^{(0,0)}$ is one of the following sets : 
\item $A_2=\{\a1\ax1-\lambda,\ldots,\an\anx-\lambda\}$
\item $A_2\cup B_1=\{\a1\ax1-\lambda,\ldots,\an\anx-\lambda,\Ax2\A2-\ax1\a1,\ldots,\anx\an-\ax1\a1\}$
\item $A_2\cup B_2=\{\a1\ax1-\lambda,\ldots,\an\anx-\lambda,\ax1\a1-\lambda,\ldots,\anx\an-\lambda\}$
\item $A_2\cup D=\{\a1\ax1-\lambda,\ldots,\an\anx-\lambda,\sum_i\aix\ai-\mu\}$
\item $A_3=\{\ai\aix+\beta\sum_j\ajx\aj-\lambda|1\leq i\leq n\}$
\item $A_1\cup D=\{\A2\Ax2-\a1\ax1,\ldots,\an\anx-\a1\ax1,\sum_i\aix\ai-\lambda\}$
\item $C=\{\sum_i\ai\aix-\lambda\}$
\item $C\cup D=\{\sum_i\ai\aix-\lambda,\sum_i\aix\ai-\mu\}$
\item $E_2=\{\ai\aix+\beta_1\aix\ai+\beta_2{\sum_{j\not=i}}\ajx\aj-\lambda|1\leq i\leq n\},\beta_1\not=\beta_2,\beta_1+(n-1)\beta_2\not=0$ or 
$\{\aix\ai+\alpha_1\ai\aix+\alpha_2{\sum_{j\not=i}}\aj\ajx-\mu|1\leq i\leq n\},\alpha_1\not=\alpha_2,\alpha_1+(n-1)\alpha_2\not=0$
\item $E_2'=\{\ai\aix+\beta_1\aix\ai+{{\beta_1\over 1-n}\sum_{j\not=i}}\ajx\aj-\lambda|1\leq i\leq n\},\beta_1\not=0$ or\newline
$\{{{\sum_i}}\ai\aix-\lambda,\alpha(\ai\aix-\a1\ax1)+\beta(\aix\ai-\ax1\a1)|i>1\}$
\item $E_1\cup C\cup
D=\{\sum_j\aj\ajx-\lambda,\sum_j\ajx\aj-\mu,\alpha\ai\aix+\beta\aix\ai-\nu|i>1\}$, with $\nu=(\alpha\lambda+\beta\mu)/n$
\item $F=\{\alpha\sum_i\ai\aix+\beta\sum_i\aix\ai-\lambda\}$
\item $A_1\cup B_1\cup F=\{\ai\aix-\a1\ax1,\aix\ai-\ax1\a1,\alpha\a1\ax1+\beta\ax1\a1-\lambda|i>1\}$
and also : $B_2$, $A_1\cup B_2$, $B_2\cup C$, $D$, $B_3$, $B_1\cup
C$, or $E_2''$, which are respectively symmetrical to $A_2$, $A_2\cup B_1$, $A_2\cup D$, $C$,
$A_3$, $A_1\cup D$, $E_2'$, by the exchange of $\ai$ and $\aix$. Moreover, in each case, at least one of the constants $\lambda$, $\mu$ or $\nu$ is non-zero, $\alpha$ and $\beta$ are non-zero, and all the constants belong to $ R$.

Such a presentation is unique, except in the cases $E_2$, $E_2'$ and $E_2''$ for which we give two forms, and is called ``standard''.
\end{proposition}

Even if a standard presentation is given and if the hypothesis $(P_4)$ is satisfied, the standard presentation could happen to be non-confluent for any ordering. The next proposition shows that it is not so. Indeed, if a confluent reduction system exists for some deglex-ordering $<$, then this system is associated with a standard presentation. Furthermore, all orderings are not allowed.

\begin{proposition}\label{prop8}
Let $I$ be an ideal such that $(P_0)$,\ldots,$(P_4)$ hold, and let $S$ be a quadratic confluent reduction system, which is adapted to some deglex ordering $<$, and associated with $I$. We also assume that $S$ is simplified and reduced. Let $R=R^{(2,0)}\coprod R^{(1,1)}\coprod\penalty -100 
R^{(1,-1)}$ $\coprod R^{(0,0)}$ be a standard presentation of $I$, and 
$R_S$ the presentation associated with $S$ and $<$.\newline
Then,
$R_S=R_S^{(2,0)}\coprod R_S^{(1,1)}\coprod R_S^{(1,-1)}\coprod
R_S^{(0,0)}$, $S=S^{(2,0)}\coprod S^{(1,1)}\coprod S^{(1,-1)}\coprod$ 
$S^{(0,0)}$, where $S^{(i,j)}$ is the reduction system associated with $R^{(i,j)}$, and $R_S^{(2,0)}=R^{(2,0)}$,
$R_S^{(1,1)}=\{x/{\rm lc}(x)|x\in R^{(1,1)}\}$, $R_S^{(1,-1)}=\{x/{\rm
  lc}(x)|x\in R^{(1,-1)}\}$. Moreover, depending on $R^{(0,0)}$, the $S^{(0,0)}$ part is :

\item{$A_2$} : $S^{(0,0)}=\{\ai\aix\r\lambda|1\leq i\leq n\}$.
\item{$A_1\cup B_2$} : $\exists i_0$ s.t. $S^{(0,0)}=\{\ai\aix\r\aiz\aizx,\ajx\aj\r\mu|i\not=i_0,1\leq j\leq n\}$
\item{$A_2\cup B_2$} : $S^{(0,0)}=\{\ai\aix\r\lambda,\aix\ai\r\mu\}$.
\item{$A_3$} : there are two possibilities (a) : $S^{(0,0)}=\{\ai\aix\r\lambda-\beta{\sum_j}\ajx\aj)\}$, or\newline
(b) : $\exists i_0,j_0$ s.t. $S^{(0,0)}=\{\ai\aix\r\aiz\aizx,\ajzx\ajz\r{\lambda\over\beta}-{1\over\beta}\aiz\aizx-{\sum_{j\not=j_0}}\ajx\aj|i\not=i_0\}$
\item{$C$} : $\exists i_0$
                  s.t. $S^{(0,0)}=\{\aiz\aizx\r\lambda-{\sum_{i\not=i_0}}\ai\aix\}$
\item{$C\cup D$} : $\exists i_0,j_0$
                  s.t. $S^{(0,0)}=\{\aiz\aizx\r\lambda-{\sum_{i\not=i_0}}\ai\aix,\ajzx\ajz\r\mu-{\sum_{j\not=j_0}}\ajx\aj\}$
\item{$A_1\cup D$} : $\exists i_0,j_0$ s.t. $S^{(0,0)}=\{\ai\aix\r\aiz\aizx,\ajzx\ajz\r\mu-{\sum_{j\not=j_0}}\ajx\aj|i\not=i_0\}$
\item{$A_2\cup D$} : $\exists i_0$ s.t. $S^{(0,0)}=\{\ai\aix\r\lambda,\aizx\aiz\r\mu-{\sum_{j\not=i_0}}\ajx\aj|1\leq i\leq n\}$ 
\item{$E_2$, with $n\geq 3$} : (a) $\{\ai\aix\r\lambda-\beta_1\aix\ai-\beta_2{\sum_{j\not=i}}\ajx\aj|1\leq i\leq n\}$, or\newline
(b) $\{\aix\ai\r\mu-\alpha_1\ai\aix-\alpha_2{\sum_{j\not=i}}\aj\ajx|1\leq i\leq n\}$
\item{$E_2$ with $n=2$} : (a), or (b) as in the previous case, or\newline
(c) $\{\ajx\aj\r{1\over\beta_2}(\lambda-\beta_1\aix\ai-\ai\aix),\aj\ajx\r{{\lambda(1-{\beta_1\over\beta_2})+({\beta_1^2\over\beta_2}-\beta_2)\aix\ai+{\beta_1\over\beta_2}\ai\aix}}\}$, with $\{i,j\}=\{1,2\}$, or\newline
(d) $\{\aix\ai\r{{1\over\beta_1}}(\lambda-\beta_2\ajx\aj-\ai\aix),\aj\ajx\r{{\lambda(1-{\beta_2\over\beta_1})+({\beta_2^2\over\beta_1}-\beta_1)\ajx\aj+{\beta_2\over\beta_1}\ai\aix}}\}$, with $\{i,j\}=\{1,2\}$
                                      
\item{$E_2'$} : (a) $\{\ai\aix\r\lambda-\beta_1\aix\ai+{{\beta_1\over n-1}\sum_{j\not=i}}\ajx\aj|1\leq i\leq n\}$, or\newline
(b)  $\{\aiz\aizx\r\lambda-{\sum_{j\not=i_0}}\aj\ajx,\aix\ai\r-\alpha\lambda+\aizx\aiz+2\alpha\ai\aix+\alpha{\sum_{j\not=i,i_0}}\aj\ajx|i\not=i_0\}$, for $i_0\in[1..n]$, or\newline 
(c) $\{\aiz\aizx\r\lambda-{\sum_{j\not=i_0}}\aj\ajx,\aizx\aiz\r-\alpha\lambda+\ajzx\ajz+2\alpha\ajz\ajzx+\alpha{\sum_{j\not=i_0,j_0}}\aj\ajx,\aix\ai\penalty-100\r\ajzx\ajz+\alpha\ajz\ajzx-\alpha\ai\aix|i\not=i_0,j_0\}$ for some $i_0$, and with $\alpha=(1-n)/\beta_1$
\item{$F$} : (a) $\{\aiz\aizx\r{{\lambda\over\alpha}}-{\sum_{i\not=i_0}}\ai\aix-{{\beta\over\alpha}\sum_{1\leq j\leq n}}\ajx\aj\}$, for some $i_0$, or\newline
(b) $\{\ajzx\ajz\r{{\lambda\over\beta}}-{{\alpha\over\beta}\sum_{1\leq i\leq n}}\ai\aix-{\sum_{j\not=j_0}}\ajx\aj\}$ for some $j_0$.
\item{$A_1\cup B_1\cup F$} : (a) $\{\ai\aix\r\aiz\aizx,\ajx\aj\r{\lambda\over\beta}-{\alpha\over\beta}\aiz\aizx|i\not=i_0,1\leq j\leq n\}$, or\newline
(b) $\{\aix\ai\r\aizx\aiz,\aj\ajx\r{\lambda\over\alpha}-{\beta\over\alpha}\aizx\aiz|i\not=i_0,1\leq j\leq n\}$
\item{$E_1\cup C\cup D$} :  $\exists{\cal
                    K},{\cal L}\subset [1,..,n]$, ${\cal K}\cap{\cal
                    L}=\emptyset$, $i_0\notin {\cal K}\cup {\cal L}$,
                    ${\cal K}\cup{\cal L}\cup\{i_0\}=[1..n]$
                    such that $S^{(0,0)}=\{\ai\aix\r{
                    {\nu\over\alpha}-{\beta\over\alpha}\aix\ai},\ajx\aj\r{ {\nu\over\beta}-{\alpha\over\beta}\aj\ajx},\aiz\aizx\r\lambda'-{\sum_{\cal K}}\ak\akx+{{\beta\over\alpha}\sum_{\cal L}}\akx\ak,\aizx\aiz\r\mu'-{\sum_{\cal L}}\akx\ak+{{\alpha\over\beta}\sum_{\cal K}}\ak\akx|i\in{\cal L},j\in{\cal K}\}$, $\lambda'=\lambda-$Card$({\cal L}){{\nu\over\alpha}}$, $\mu'=\mu-$Card$({\cal K}){{\nu\over\beta}}$.
 
By the exchange of $\ai$ and $\aix$ we also find the forms corresponding to $B_2$, $A_2\cup B_1$, $B_3$, and $E_2''$.
\end{proposition}

For the proof, see \cite{bes1}.

The next proposition will allow us to use symmetry between $\ai$ and $\aix$ and fix the value of some constant when needed.
  
\begin{proposition}\label{prop9}
Let $\phi$ be one of the following automorphisms of $L_n$ :

\hfil $\matrix{\epsilon : L_n & \rightarrow & L_n \cr
                            \ai & \longmapsto & \aix \cr
                            \aix & \longmapsto & \ai \cr
}$\hfil$\matrix{\phi_{\lambda,\mu} : L_n & \rightarrow & L_n \cr
                                \ai     & \longmapsto & \lambda\ai \cr
                                \aix    & \longmapsto & \mu\aix \cr
}$\hfil\newline
with $\lambda\mu\in  R\setminus\{0\}$. Then if $I$ fulfils $(P_0)$,\ldots,\penalty -100$(P_4)$, so does $\phi(I)$.
\end{proposition}
\dem\newline
We do it for $\phi_{\lambda,\mu}$ the case of $\epsilon$ is even easier and is left to the reader.

For $(P_0)$, it is obvious. Let us show that $\phi_{\lambda,\mu}(I)$ fulfils $(P_1)$ : it is only needed to verify that the image under $\phi_{\lambda,\mu}$ of a standard presentation of $I$, which is a presentation of $\phi_{\lambda,\mu}(I)$, is sent into $\phi_{\lambda,\mu}(I)$ by $J$.

If $x$ is a homogenous quadratic element of a standard presentation of $I$, $\phi_{\lambda,\mu}(x)$ is
proportional to $x$, so $J(\phi_{\lambda,\mu}(x))$ is proportional to $\phi_{\lambda,\mu}(J(x))$ and therefore belongs to $\phi_{\lambda,\mu}(I)$. Now if $x=\sum\alpha_i\ai\aix+\sum\beta_i\aix\ai-\nu.1$, with $\alpha_i$,
$\beta_i$, $\nu$ $\in R$, is a generator of $I$, then $\phi_{\lambda,\mu}(x)=\lambda\mu(\sum\alpha_i\ai\aix+\sum\beta_i\aix\ai)-\nu.1$ is stable under $J$, because $\lambda\mu\in R$.

It is clear by its definition that $\phi_{\lambda,\mu}$ commutes with the action of $\SS_n$. Thus, $\sigma^*(\phi_{\lambda,\mu}(I))=\phi_{\lambda,\mu}(\sigma^*(I))=\phi_{\lambda,\mu}(I)$

Lastly, let $S$, whose elements we denote by $m_s\r f_s$, be a quadratic confluent reduction system, adapted to $<$ and associated with $I$, and let $W$ be the vector space spanned by the irreducible monomials relatively to $S$. If we set $\phi_{\lambda,\mu}(S)=\{m_s\r{
  {1\over\lambda^k\mu^l}}\phi_{\lambda,\mu}(f_s)|s\in S,$ $m_s$ of degree $k$ in $\ai$ and $l$ in $\aix\}$, then $\phi_{\lambda,\mu}(W)=W$ is also the linear span of monomials that are irreducible under $\phi_{\lambda,\mu}(S)$. Moreover $\phi_{\lambda,\mu}(S)$ is clearly adapted to $<$ and $I\oplus W=L_n\Rightarrow\phi_{\lambda,\mu}(I)\oplus\phi_{\lambda,\mu}(W)=L_n$. Thus, by Bergman's lemma, $\phi_{\lambda,\mu}(S)$ is confluent.\hfill QED.
\smallbreak
The last of our lemmas will help us to reduce the number of cases.

\begin{lemma}\label{lem12}
If $I$ fulfils $(P_0)$,\dots,$(P_4)$, and if $n\geq 2$, then $I$ must contain a set of generators of type (1,1) or (1,-1). 
\end{lemma} 
\dem\newline
Let $S$ be a quadratic confluent reduction system for $I$, associated with some deglex-ordering $<$, $R$ be the associated presentation, $T$ the basis of irreducible monomials, and $W=$ Span$(T)$. Let us denote by $\tilde{N}_1,\ldots,\tilde{N}_n$ the representatives of $N_1,\ldots,N_n$ in $W$, and write $\tilde{N}_1=\lambda_1\xi_1+\ldots+\lambda_k\xi_k+u$,
with $d^\circ(\xi_1)=\ldots=d^\circ(\xi_k)>d^\circ u$ and $\xi_1>\ldots>\xi_k$,
$\lambda_1,\ldots,\lambda_k\in K\setminus\{0\}$.

We first need to show the following formula : 
\be
\forall i,j,\quad {\cal N}_i\tilde{N}_j=0\label{for1}
\ee
This is true because $[N_i,N_j]=0\ssi{\cal N}_i\tilde{N}_j\in I$. But $T$ is made of eigenvectors for ${\cal N}_i$, so $W$ is stable under ${\cal
  N}_i$, consequently ${\cal N}_i\tilde{N}_j\in W\cap I=\{0\}$.

Obviously, we also have :
\be
\forall i,j,\quad {\cal N}_i\xi_j=0\label{for2}
\ee
Write $\xi_1=x\eta y$, with $x,y\in X$. There are two cases :
\begin{itemize}
\item{} There exists $b\in X$ with an index different from the indices of $x$ and $y$. Suppose first that $b>x$. Then $b\xi_1>\xi_1b$, and since $b\xi_1>b\xi_2>\ldots$ and $\xi_1b>\xi_2b>\ldots$ we have lm$([\tilde N_1,b])=b\xi_1$. Thus $b\xi_1=bx\eta y$ must be reducible, but since $\xi_1=x\eta y$ is not and the reduction system is quadratic, it is only possible if $bx$ is reducible. We then have relations of type $(1,1)$ or $(1,-1)$ in $R$. Now if $b<x$, we have lm$([\tilde N_1,b])=\xi_1b=x\eta yb$, and we deduce that $yb$ is reducible. Therefore the relations $(1,1)$ or $(1,-1)$ are in $R$.
\item{} There is no such $b$. Then $n=2$ and $X=\{x,x^+,y,y^+\}$. If $xy$ or $yx$ are reducible we have relations $(1,1)$ or $(1,-1)$, so we suppose they are irreducible. For $k$ large enough, we have $(xy)^k\xi_1\not=\xi_1(xy)^k$. Indeed, if not we would conclude that $\xi_1$ divides $(xy)^k$ but it is impossible since $\xi_1$ must contain $x^+$ or $y^+$ by the formula (\ref{for2}). So we deduce that lm$([\tilde N_1,(xy)^k])=\xi_1(xy)^k$ or $(xy)^k\xi_1$. In both cases it is irreducible, so this is a contradiction.\hfill QED.
\end{itemize}

\subsection{The main calculations}

If $I$ is such that $(P_0),\ldots,(P_4)$ hold, it has a standard presentation by proposition \ref{prop7}. We have to study every such presentation that is not yet ruled out by lemma \ref{lem12}. In most cases, it is possible to show that $(P_0)$, $(P_3)$, or $(P_4)$ cannot hold by making use of our different lemmas. Nonetheless, it is sometimes necessary to call upon a reduction system and calculate in a basis of irreducible monomials.

For convenience, we will deal with relations in $B$ rather than with generators of the ideal $I$.

When the relations $(0,0)$ depend on a single constant term $\lambda$ (which must be non-zero) we assume that $\lambda=1$.

In cases containing $(1,-1)_a$, and if $r\not=0$, we set ${\displaystyle q=-{s\over r}}$. If $s\not=0$, we set ${\displaystyle q'=-{r\over s}}$.

From now on we assume $n\geq 2$. We will look at the case $n=1$ afterwards.
\smallbreak

\underline{$(1,1)_c\cup (1,-1)_b\cup (2,0)$ :}\newline
So $I$ is generated by the relations $(1,1)_c\cup (1,-1)_b\cup (2,0)$ together with some relations of type $(0,0)$. If relations of type $A_2$, $C$, $B_2$ or $D$ with $\lambda$ or $\mu\not=0$ are present, it is easily seen by multiplying them on the left or on the right by $\ai$ that $\ai=0$. Thus $(P_0)$ is not satisfied. Let us see the other cases :
        \begin{itemize}
        \item{$A_3$} : In $B$ we have :
        $$\left\{\matrix{
        \a1\ax1+\beta\sum\aix\ai=1 & (l_1) \cr
        \ldots & \cr
        \an\anx+\beta\sum\aix\ai=1 & (l_n) \cr}\right.$$
	Let's multiply $(l_1)$ on the right by $\an$, we get 
        $\an=0$, thus $B=0$.           
        This also rules out the case $A_1\cup B_1\cup
        F=A_3\cup B_1$.

        \item{$E_2$} : 
          $$\left\{
        \matrix{
        \a1\ax1+\beta_1\ax1\a1+\beta_2\sum_{i\not=1}\aix\ai=1 & (l_1) \cr
        \ldots  &  \cr
        \an\anx+\beta_1\anx\an+\beta_2\sum_{i\not=n}\aix\ai=1 & (l_n) \cr
        }
        \right.$$
	We do as above.

        \item{$F$} : 
          $$\alpha\sum\ai\aix+\beta\sum\aix\ai=1$$
 We first multiply the relation $F$ by $\ai$ on the left, then on the right, and we get :
        $$\beta\ai\aix\ai=\alpha\ai\aix\ai=\ai$$
        If $\alpha\not=\beta$, we have $B=0$. Thus we can assume that $\alpha=\beta=1$.

        We have $\ai\aix\ai=\ai$, $\forall i$. But if there exists a quadratic confluent reduction system for $I$, it is of type (a) or (b) (see proposition 2). In both cases, there exists $i$ such that $\ai\aix$ and 
        $\aix\ai$ are irreducible, so
        $\ai\aix\ai$ must be irreducible, and this is a contradiction.
        \end{itemize}
   
\smallbreak

\underline{$(1,1)_{a\mbox{ or }b}\cup (1,-1)_b\cup (2,0)$ :}\newline
Making use of either $(1,-1)_{a,b}$ or $(2,0)$, we see that $A_2$ or $B_2$
with $\lambda\not=0$ leads to $B=0$.
        \begin{itemize}
        \item{$C$} : Multiplying the relation $(C)$ by $\a1$ on the left we get :
        $$ \a1^2\ax1+\sum_{i>1}\a1\ai\aix=\a1$$
        $$\Rightarrow \sum_{i>1}\pm\ai\a1\aix=\a1$$
        $$\Rightarrow 0=\a1\Rightarrow B=0$$
	This also rules out all other presentations containing relations of type $C$ (in particular $E_2'=E_1\cup C$).

        \item{$F$} : We have
        $$\alpha\sum_{1\leq i\leq n}\a1\ai\aix+\beta\sum_{1\leq i\leq n}\a1\aix\ai=\a1$$
        $$\Rightarrow \alpha\sum_{i>1}\pm\ai\a1\aix+\beta\a1\ax1\a1=\a1$$
        $$\Rightarrow \beta\a1\ax1\a1=\a1$$
	Multiplying by $\a1$ on the right we would obtain in the same way :
\be
\alpha\a1\ax1\a1=\a1\label{eq3}
\ee
        thus $\alpha=\beta$. Furthermore, multiplying (\ref{eq3}) by $\ai$, $i\not=1$ we get :
        $$\a1\ai=\alpha\a1\ax1\a1\ai=\pm\alpha\a1\ax1\ai\a1=0$$
        But, the presentation being standard, 
        $\{\ai\aj-\aj\ai,\aix\ajx-\ajx\aix|i<j\}$ is a basis 
        of $I_2^{(1,1)}$. Now 
        $\a1\ai\notin$ Span$\{\ai\aj-\aj\ai,\aix\ajx-\ajx\aix\}$, a contradiction.\newline
        We can do the same for $F$, and thus for $A_3$, $B_3$ and $E_2$.
        \end{itemize}
\smallbreak

\underline{$(1,1)_c\cup(1,-1)_a\cup(2,0)$} :
        \begin{itemize}
        \item{$A_3$} :
        Let's multiply $(l_1)$ on the right by $\a1$ :
\be      
\a1\ax1\a1=\a1\label{eq4}
\ee
Now $\a1\ax1=\ai\aix\Rightarrow\a1=\ai\aix\a1$, $\forall i$. Then if
        $s\not=0$, we have $\a1=-q'\ai\a1\aix=0$,
        for $i\not=1$, and $B=0$. 
	We can thus assume that $s=0$.\newline
        Now $\ai\ajx=0$, and multiplying $(l_1)$ to the left 
        by $\a1$, we find $\a1=\beta\a1\ax1\a1$, consequently we have $\beta=1$, by (\ref{eq4}).\newline
	We are then in the case (b) of the theorem. It is easily seen that the reduction system 
$\{\ai\aj\r 0,\aix\ajx\r 0,\ai^2\r 0,\aix^2\r 0,\ai\ajx\r 0,\ai\aix\r 1-\sum\akx\ak|1\leq i\not=j\leq n\}$ is confluent and adapted to the deglex-ordering coming from $\ax1<\ldots<\anx<\a1<\ldots<\an$, which we will denote by $<_n$ in the rest of the section. We let the reader verify that $(P_3)$ holds, with $N_i=\aix\ai+\lambda_i.1$, $\lambda_i\in K$.

\item{$F$} : Let's multiply $F$ by $\a1$ on the right, then on the left :
$$\left\{\matrix{
\beta\a1(\sum_{1\leq i\leq n}\aix\ai)=\a1\cr
\alpha(\sum_{1\leq i\leq n}\ai\aix)\a1=\a1\cr}\right.$$
$$\Rightarrow\left\{\matrix{r\a1=r\beta\a1\ax1\a1+\beta\sum_{i>1}(-s\aix\a1)\ai\cr
s\a1=s\alpha\a1\ax1\a1+\alpha\sum_{i>1}\ai(-r\a1\aix)\cr}\right.$$
$$\Rightarrow\left\{\matrix{r\a1=r\beta\a1\ax1\a1\cr
s\a1=s\alpha\a1\ax1\a1\cr}\right.$$
$$\Rightarrow\left\{\matrix{r\a1\Ax2=-s\beta\a1\ax1\Ax2\a1=0\cr
s\Ax2\a1=-r\alpha\a1\Ax2\ax1\a1=0\cr}\right.$$
Thus $r=0$ or $s=0$ (the presentation is standard). If
$r=0$, $\ajx(\alpha\sum_i\ai\aix)+\ajx\beta(\sum\aix\ai)=\alpha\ajx\aj\ajx=\ajx$.
Now $\exists j$ such that $\ajx\aj\ajx$ is irreducible, and we come to a contradiction. The case $s=0$ is symmetrical.

\item{$A_1\cup B_1\cup F$} : From case $F$ we know that we can assume that $r=0$, then we have $\an=\alpha\a1\ax1\an+\beta\ax1\a1\an=0$. Consequently $B=0$.

\item{$E_2$} : By $F$, we have $rs=0$. In each case it is easy to show that $B=0$.
\end{itemize}
\smallbreak

\underline{$(1,1)_b\cup(1,-1)_a\cup(2,0)$} :
\begin{itemize}
\item{$C$} :
Let's show that $B=0$ :
$$\sum\ai\aix=1$$
$$\Rightarrow\sum\a1\ldots\an\ai\aix=\a1\ldots\an$$
$$\Rightarrow\sum\a1\ldots a_{i-1}^{}a_{i+1}^{}\ldots\an\ai\ai\aix=\a1\ldots\an$$
$$\Rightarrow 0=\a1\ldots\an$$
Suppose that every product $a_{i_1}\ldots a_{i_k}$
of length $k$ is zero. This is true for $k=n$. Then : 
$$a_{i_1}^{}\ldots a_{i_{k-1}}^{}=\sum a_{i_1}^{}\ldots
a_{i_{k-1}}^{}\ai\aix$$
and the sum is zero, since all terms $a_{i_1}^{}\ldots a_{i_{k-1}}^{}\ai$ vanish. Thus, by induction, we see that 
$B=0$.

Since we made no use of relations $(1,-1)_a$, we can get rid of the case $(1,1)_b\cup(2,0)\cup C$ by the same method.

\item{$F$} : Thanks to all the relations we have, one can show that

$$(\alpha-\beta)(r+s)\a1\ldots\an=0$$
Then, if $(\alpha-\beta)(r+s)\not=0$, it can be proved by induction that $B=0$. If only one of the factors $\alpha-\beta$ or $r+s$ vanishes then it is possible to show that $\a1\an=0$. Let's see the case $\alpha=\beta$ and $r+s=0$ in more details. We can assume $\alpha=1$.

If $n\geq 2$, it is easily verified that the natural projection $L_n\rightarrow L_2$ gives a surjective homomorphism from $B_n$ onto $B_2$. Therefore, if $(P_3)$ does not hold for $n=2$, it will not hold for any $n\geq 2$.

Let $<$ be the only deglex-ordering such that $\ax1<\a1<\Ax2<\A2$. The reduction system 
$S=\{\A2\a1\r \a1\A2,\Ax2\ax1\r\ax1\Ax2,\a1^2\r 0,\ax1^2\r 0,\A2^2\r 0,$ $\Ax2^2\r 0,\A2\ax1\r\ax1\A2,\Ax2\a1\r\a1\Ax2,\A2\Ax2\r 1-\a1\ax1-\Ax2\A2-\ax1\a1\}$ is confluent and adapted to $<$. If we call $T$ the basis of irreducible monomials and $T_0:=T\cap
B_2^0$, then it is clear that $T_0=\{1,\Ax2\A2,(\ax1\a1)^k,\penalty -1000(\ax1\a1)^k\Ax2\A2,(\a1\ax1)^k,(\a1\ax1)^k\Ax2\A2)|k\geq 1\}$.
Indeed, no $\A2$ can be on the left of another term, an $\Ax2$ can only be on the left of an $\A2$, etc\ldots Now, for $k\geq 1$ : $[(\ax1\a1)^k,\ax1]=(\ax1\a1)^k\ax1$,
$[(\ax1\a1)^k\Ax2\A2,\ax1]=(\ax1\a1)^k\ax1\Ax2\A2$, $[(\a1\ax1)^k\Ax2\A2,\ax1]=-\ax1(\a1\ax1)^k\Ax2\A2$, $[(\a1\ax1)^k,\ax1]=-\ax1(\a1\ax1)^k$ and $[\Ax2\A2,\ax1]=0$.\newline
We see that $\ax1$ never appears in these commutators. Consequently, $(P_3)$ cannot hold.

\item{$A_3$} : From $F$ we have $r+s=0$, then :
$$\matrix{
\a1\ldots\an & =
&\a1\ax1\a1\ldots\an+\beta\sum_{i\not=1}\aix\ai\a1\ldots\an \hfill\cr
              & = & \a1\ax1\a1\ldots\an\hfill \cr
              & = & \A2\Ax2\a1\ldots\an \hfill\cr
              & = & \a1\A2\Ax2\A2\ldots\an\hfill \cr
              & = & \a1(\a1\ax1)\A2\ldots\an\hfill \cr
              & = & 0\hfill\cr
}$$
So by induction $B=0$.

\item{$E_2$} : From $F$, we can assume that $r+s=0$ and
  $\beta_1+(n-1)\beta_2=n$ (which corresponds to $\beta=1$ in the case $F$).\newline
$\a1(l_1)-(l_1)\a1$ gives :
$$(\beta_1-1)\a1\ax1\a1+\beta_2\sum_{i\not=1}(\a1\aix-\aix\a1)\ai=0$$

\be
\Rightarrow (\beta_1-1)\a1\ax1\a1=0\label{star1}
\ee

But we also have :
$$\displaylines{
\a1(l_n)-(l_n)\a1=\a1\an\anx+\beta_1\a1\anx\an+\beta_2\sum_{1<i<n}\a1\aix\ai-\an\anx\a1\hfill\cr
\hfill +\beta_2\a1\ax1\a1-\beta_1\anx\an\a1-\beta_2\sum_{1<i<n}\aix\a1\ai\cr}
$$
\be
\Rightarrow 0=\beta_2\a1\ax1\a1\label{star2}
\ee
        \begin{itemize}
        \item{if $\beta_2\not=0$}, $(\ref{star2})\Rightarrow\a1\ax1\a1=0$. Thanks to $\a1\ldots\an(l_i)$ we get $\a1\ldots\an=0$ and by induction $B=0$.
        \item{if $\beta_2=0$}, we can assume $\beta_1=1$, or else
          $B=0$ by (\ref{star1}). We get the reduction system $\{\ai^2\r 0,\aix^2\r 0,\aj\ai\r\ai\aj,\ajx\aix\r\aix\ajx,\ai\ajx\r$ $\ajx\ai,\ai\aix\r 1-\aix\ai|1\leq i<j\leq n\}$. It is confluent and adapted to $<_n$. $B$ is a solution to our problem, with $N_i=\aix\ai+\lambda_i.1$,
        $\lambda_i\in K$, and we are in case $(a)$ of theorem 1.
        \end{itemize}

\end{itemize}
\smallbreak

\underline{$(1,1)_a\cup(1,-1)_a\cup(2,0)$} : \newline
This case is similar to the preceding one. In the case $E_2$, we find the solution $(a')$.
\smallbreak

\underline{$(1,1)_c\cup(1,-1)_b$} :

\begin{itemize}

\item{$C$} : Let $\phi$ : $L_n\rightarrow C$, with 
${\displaystyle C=\bigoplus_{i=1}^n K[x_i,y_i]/\langle x_iy_i-1\rangle }$, defined by :\penalty-1000 $\phi(\ai)=x_i$, $\phi(\aix)=y_i$. $\phi$ goes to the quotient, indeed : $\forall i\not=j$, $\phi(\ai\aj)=x_ix_j=0$,
$\phi(\aix\ajx)=y_iy_j=0$, $\phi(\ai\ajx)=\phi(\ajx\ai)=x_iy_j=0$,
$\phi(\sum\ai\aix-1)=(\sum x_iy_i)-1=0$. But $\phi\not=0$ then by lemma \ref{lem4}, $(P_3)$ does not hold.

\item{$C\cup D$} : We have $(C)\a1\Rightarrow \a1\ax1\a1=\lambda\a1$ and $\a1(D)\Rightarrow \a1\ax1\a1=\mu\a1$, then if $\lambda\not=\mu$, $B=0$. If $\lambda=\mu$, we do as above, using the same $\phi$.

\item{$A_2$, $\lambda=0$, $\cup D$} : Computing $(D)\a1$ we are led to $\a1^2=0$.

\item{$A_3$ or $A_1\cup D$} : With $\alpha=1$ or $0$ : 
$$\alpha\a1\ax1+\beta\sum\aix\ai=1$$
$$\Rightarrow \alpha\a1^2\ax1+\beta\a1\ax1\a1=\a1$$
$$\Rightarrow \alpha\a1\A2\Ax2+\beta\a1\ax1\a1=\a1$$
$$\Rightarrow \beta\a1\ax1\a1\ax1=\a1\ax1$$
$$\Rightarrow \beta\a1\ax1\A2\Ax2=0=\a1\ax1$$
This is impossible in both cases. We can do the same for 
$A_1\cup B_1\cup F$.

\item{$E_2$ or $E_2'$} : Let's calculate $\a1(l_1)$, $(l_1)\a1$, $\a1(l_n)$, and $(l_n)\a1$. We get 
:
$$\left\{\matrix{
\a1^2\ax1+\beta_1\a1\ax1\a1=\a1 & (i)\cr
\a1\ax1\a1+\beta_1\ax1\a1^2=\a1 & (ii)\cr
\beta_2\a1\ax1\a1=\a1 & (iii)\cr
\beta_2\ax1\a1^2=\a1 & (iv)\cr
}\right.$$
Thanks to the three last formulas we get 
$(\beta_2-\beta_1-1)\a1=0$. 

So, if $\beta_2-\beta_1-1\not=0$, i.e. ${\beta_1\not={n-1\over n}}$ in the case $E_2'$, $B=0$.\newline
Moreover, $(iii)$ and $(iv)$ $\Rightarrow \beta_2\not=0$, or else 
$B=0$.\newline
The only remaining case is $\beta_2=\beta_1+1\not=0$. We define $\phi : B\rightarrow {\displaystyle
  \bigoplus_{i=1}^nK[x_i,y_i]/\langle x_iy_i-1/\beta_2\rangle}$ by 
$\phi(\ai)=x_i$, $\phi(\aix)=y_i$. $\phi$ is well defined since 
$\phi(\ai\aix+\beta_1\aix\ai+\beta_2\sum_{j\not=i}\ajx\aj)=(1+\beta_1)x_iy_i+\beta_2\sum_{j\not=i}
x_jy_j=1$. Thus $(P_3)$ cannot hold.

\item{$E_1\cup C\cup D$} :
$$\left\{\matrix{
\a1\ax1+\ldots+\an\anx=\lambda & (l_1)\cr
\ax1\a1+\ldots+\anx\an=\mu & (l_2)\cr
\alpha\ai\aix+\beta\aix\ai=\nu & (l_{3,i})\cr
}\right.$$
We must have $\lambda=\mu$ by $C\cup D$. Now, if we multiply $(l_{3,i})$ by $\aj$ with $j\not=i$, we find 
$\nu\aj=0$, therefore $\nu=0$. But 
$\lambda(\alpha+\beta)=n\nu=0$ and $\lambda$ must be non-zero by lemma \ref{lem3}, then $\alpha+\beta=0$. We can assume that 
$\lambda=1$ and $\alpha=-\beta=1$. We then use the same $\phi$ as in case $E_2$.

\item{$F$} :
	 $$\alpha\sum \ai\aix-\beta\sum\aix\ai=1$$
	$$\Rightarrow \alpha\ai^2\aix-\beta\ai\aix\ai=\ai$$

                Now $(P_4)$ implies that there must always exist $i$ s.t. $\ai\aix$ and $\aix\ai$ are irreducible. A contradiction.\newline

\end{itemize}

\smallbreak

\underline{$(1,1)_{a\mbox{ or }b}\cup(1,-1)_b$} : Such a presentation is never standard. Indeed, we always have relations of the form :
$$\sum_{1\leq i\leq n}\alpha_i\ai\aix+\sum_{1\leq i\leq
  n}\beta_i\aix\ai+\lambda=0$$
with $\lambda\not=0$. Therefore :
$$\a1\sum_{1\leq i\leq
  n}\alpha_i\ai\aix+\beta_1\a1\ax1\a1+\lambda\a1=0$$
$$\Rightarrow
\alpha_2\a1\A2\Ax2\A2+\beta_1\a1\ax1\a1\A2+\lambda\a1\A2=0$$
$$\Rightarrow
\pm\alpha_2\A2(\a1\Ax2)\A2\pm\beta_1\a1(\ax1\A2)\A2+\lambda\a1\A2=0$$
$$\Rightarrow \lambda\a1\A2=0$$
$$\Rightarrow \a1\A2=0$$
This is impossible.

\smallbreak

\underline{$(1,1)_c\cup(1,-1)_a$} :
This case is very easy, and we only state the results.
        \begin{itemize}
        \item{$C$, $C\cup D$, $A_1\cup D$} : Multiplying the relation $C$ or $D$ by $\ai$, one can prove that the presentation is not standard.
        \item{$A_3$ or $A_1\cup B_1\cup F$} : It can be proven that $\ai^2=0$, thus the presentation is not standard.
        \item{$E_2$ or $E_2'$} : It can be shown that :
        $$ rs(\beta_2-\beta_1-1)\a1=0$$
        Therefore we have $rs(\beta_2-\beta_1-1)=0$. If $\beta_2-\beta_1-1=0$ the homomorphism $\phi$ of $(1,1)_c\cup (1,-1)_b\cup E_2$ can be used to exclude this case. If $rs=0$ and $\beta_2-\beta_1-1\not=0$, the presentation is shown to be non-standard.

        \item{$F$} : We have, $\forall i$ : $\alpha\ai^2\aix\ai+\beta\ai\aix\ai^2=\ai^2$, so we conclude that the presentation is not standard.
   
\end{itemize}
\smallbreak

\underline{$(1,1)_b\cup(1,-1)_a$} :
        \begin{itemize}
        \item{$A_2$} : 
        $$r\A2=r\A2\a1\ax1=r\a1\A2\ax1$$
        $$\Rightarrow -s\a1\ax1\A2=r\A2$$
        $$\Rightarrow (r+s)\A2=0$$
        Thus $r+s\not=0\Rightarrow B=0$. If $r+s=0$ an homomorphism 
        $\delta : B\rightarrow C$, with
        $C=K[x_1,\ldots,x_n,y_1,\ldots,y_n]/\langle x_1y_1-1,\ldots,x_ny_n-1\rangle $, is defined by setting $\delta(\ai)=x_i$, $\delta(\aix)=y_i$. Therefore $(P_3)$ cannot hold, by lemma \ref{lem4}. We do the same for $A_2\cup B_1$ and $A_2\cup B_2$.

        \item{$A_3$, $A_1\cup D$ and $A_2\cup D$} : In these 3 cases we have  :
          $$\left\{\matrix{\ai\aix=\aj\ajx\cr
              \alpha\ai\aix+\beta\sum_j\ajx\aj=1\cr}\right.$$
          with $\alpha=0$ for $A_2\cup D$ and $A_1\cup D$, and $\alpha=1$ for $A_3$.\newline 
         If we set 
        $x=\a1\ax1=\ldots=\an\anx$, we easily show that $\forall i$, $r\ai x+sx\ai=0$, and $\forall i\not=1$, $r\a1\aix\ai+s\aix\ai\a1=0$. Then :

        $$ \beta(r\a1\ax1\a1+s\ax1\a1^2)=(r+s)\a1$$
        $$\Rightarrow \beta(r\a1^2\ax1\a1+s\a1\ax1\a1^2)=(r+s)\a1^2$$
        $$\Rightarrow \beta(r\a1 x+sx\a1)\a1=(r+s)\a1^2$$
        $$\Rightarrow 0=(r+s)\a1^2$$
        Therefore $r+s=0$. Let's show that $x:=\ai\aix$ and $y_i:=\aix\ai$ are central elements of $B$. For $x$ it is trivial : we calculate the commutator of $x$ with $\ak$ or $\akx$ by writing $x=\ai\aix$ with $i\not=k$. Now $y_i$ clearly commutes with $\aj$ and $\ajx$ for $j\not=i$. As for $j=i$, we just need to write 
$$y_i={1\over\beta}-{1\over\beta}x-\sum_{k\not=i}\akx\ak$$
Now we show that $B^0\subset Z(B)$. Let $m\in B^0$ be a monomial, and write $m$ as $m=b_1\ldots b_{2l}$, with $b_i=\ai$ or $\aix$. There must exists $i<j$ such that $b_j=b_i^+$ ($:=J(b_i)$) and $\forall k$, $i<k<j$, $b_k\not=b_i$ and $b_k\not=b_i^+$. Then $b_i$ commutes with every $b_k$ s.t. $i<k<j$, and we can write : 
$$\matrix{m&=&b_1\ldots b_{i-1}b_ib_{i+1}\ldots b_{j-1}b_jb_{j+1}\ldots b_{2l}\hfill\cr
	&=&b_1\ldots b_{i-1}b_{i+1}\ldots b_{j-1}(b_ib_j)b_{j+1}\ldots b_{2l}\hfill\cr}$$
Now $b_ib_j=b_ib_i^+=x$ or $y_i$, thus $b_ib_i^+\in Z(B)$. Consequently $m=b_1\ldots b_{i-1}b_{i+1}\ldots b_{j-1}b_{j+1}\ldots b_{2l}(b_ib_i^+)$. Let $m=m'(b_ib_i^+)$, with $b_ib_i^+\in Z(B)$ and $m'\in B^0$. By an easy induction, we find $m\in Z(B)$. Then $B^0\subset Z(B)$ (the other inclusion is always true). As a consequence, $(P_3)$ cannot hold (unless $B=0$).

        \item{$A_1\cup B_1\cup F$} : As above, we must have $r+s=0$. If $\alpha+\beta\not=0$, lemma \ref{lem4} can be used, with the help of $\gamma : B\rightarrow K[x,y]/\langle xy-1/(\alpha+\beta)\rangle $, defined by $\gamma(\ai)=x$, $\gamma(\aix)=y$. If $\alpha+\beta=0$ we have (with $\alpha=1$):
                $$\a1\ax1\a1-\ax1\a1^2=\a1$$
                $$\Rightarrow \a1\aix\ai-\ax1\a1^2=\a1,\ i\not=1$$
                $$\Rightarrow \aix\ai\a1-\ax1\a1^2=\a1$$
                $$\Rightarrow \ax1\a1^2-\ax1\a1^2=0=\a1$$
                $$\Rightarrow B=0 $$

        \item{$C$} : We do as in case $(1,1)_c\cup(1,-1)_b\cup C$. 

        \item{$C\cup D$} : 
\be
r\a1(C)+s(C)\a1\Rightarrow
r\a1^2\ax1+s\a1\ax1\a1=\lambda(r+s)\a1\label{star3}
\ee
\be
r\a1(D)+s(D)\a1\Rightarrow
        r\a1\ax1\a1+s\ax1\a1^2=\mu(r+s)\a1\label{star4}
\ee
        From these two relations we get :
        $$\left\{\matrix{
                r\a1^2\ax1\a1+s\a1\ax1\a1^2=\lambda(r+s)\a1^2\cr
                r\a1^2\ax1\a1+s\a1\ax1\a1^2=\mu(r+s)\a1^2\cr}\right.$$
        $$\Rightarrow (r+s)(\lambda-\mu)\a1^2=0$$
        This shows that 
        $(r+s)(\lambda-\mu)=0$.
                \begin{itemize}
                \item{$\lambda=\mu$} : We can use $\phi$ as in case $C$.
                \item{$\lambda\not=\mu,r+s=0$} then (\ref{star3}) and (\ref{star4})
                $\Rightarrow \a1^2\ax1=\a1\ax1\a1=\ax1\a1^2$. Thus :
                $$ \matrix{\lambda\a1\ldots\an & = &
\a1\ldots\an\sum\ai\aix\hfill\cr
                                               & = &
\sum\ai^2\aix\a1\ldots a_{i-1}^{}a_{i+1}^{}\ldots\an\hfill\cr
                                               & = &
\sum\aix\ai(\a1\ldots\an)\hfill\cr
                                               & = &
\mu\a1\ldots\an\hfill\cr}$$
                Therefore $\a1\ldots\an=0$, and, by induction : $B=0$.
                \end{itemize}

        \item{$E_2$ or $E_2'$} : On one hand $r\a1(l_1)+s(l_1)\a1$ gives 

\be
(r+s)\a1=s\beta_1\ax1\a1^2+r\a1^2\ax1+(s+r\beta_1)\a1\ax1\a1\label{eqi}
\ee

on the other hand $r\a1(l_2)+s(l_2)\a1$ gives :
\be
s\beta_2\ax1\a1^2+r\beta_2\a1\ax1\a1=(r+s)\a1\label{eqii}
\ee
From (\ref{eqii}) we get : 
$$\beta_2=0\Rightarrow r+s=0$$ 
Furthermore :
                        $$\left\{\matrix{
                                \beta_2 (\ref{eqi})-\beta_1(\ref{eqii})\cr
                                (\ref{eqii})\cr}\right.\ssi (S) :\left\{\matrix{
                                                         \beta_2\a1(r\a1\ax1+s\ax1\a1)=(\beta_2-\beta_1)(r+s)\a1 \hfill\cr
                                                        \beta_2(r\a1\ax1+s\ax1\a1)\a1=(r+s)\a1\hfill\cr}\right.$$
                        Now :
                        $$(S)\Rightarrow \left\{\matrix{
                                                \beta_2\a1(r\a1\ax1+s\ax1\a1)\a1=(\beta_2-\beta_1)(r+s)\a1^2 \cr
                                                \beta_2\a1(r\a1\ax1+s\ax1\a1)\a1=(r+s)\a1^2\cr}\right.$$
                        $$\Rightarrow
(\beta_2-\beta_1-1)(r+s)\a1^2=0$$
                        Then we must have 
$(r+s)(\beta_2-\beta_1-1)=0$.
                                \begin{itemize}
                                 \item{$\beta_2-\beta_1-1=0$} :
                                       \begin{itemize}
                                        \item{$\beta_2\not=0$}, we define 
                                        $\rho : L_n\rightarrow C$,
                                        with $C=K[x,y]/\langle xy-{1\over
                                        \beta_2}\rangle $, such that 
                                        $\rho(\a1)=x$, $\rho(\ax1)=y$,
                                        and $\rho(\ai)=\rho(\aix)=0$,
                                        $\forall i>1$. We then see that 
                                        $\rho$ is well defined non-zero homomorphism to $C$.
                                        \item{$\beta_2=0$} (which entails $r+s=0$); $\beta_1=-1$. We then have the presentation $\{\ai\aj-\aj\ai,\aix\ajx-\ajx\aix,\ai\ajx-\ajx\ai,\ai\aix-\aix\ai-1\}$.
                                        We recognize the Weyl algebra $A_n$,
                                        which is a well known 
                                        solution to our problem, with
                                        $N_i=\aix\ai+\lambda_i.1$,
                                        $\lambda_i\in K$.
                                        \end{itemize}
                                \item{$1+\beta_1-\beta_2\not=0$,
                                $r+s=0$}
                                $$(S) \ssi\left\{\matrix{
                                        \beta_2\a1[\a1,\ax1]=0\cr
                                        \beta_2[\a1,\ax1]\a1=0\cr}\right.$$
                                where $[.,.]$ denotes the commutator.
                                        \begin{itemize}
                                        \item{$1+\beta_1+(n-1)\beta_2\not=0$} (always true in the case $E_2'$) : We can then define a homomorphism $\xi$ from $B$ to $C=K[x_1,\ldots,x_n]/\langle x_i^2-{\displaystyle {1\over 1+\beta_1+(n-1)\beta_2}}\rangle $ by setting : $\xi(\ai)=\xi(\aix)=x_i$.
                                        \item{$1+\beta_1+(n-1)\beta_2=0$}
                                        $$(l_1)+\ldots+(l_n)\ssi\sum\ai\aix+\beta_1\sum\aix\ai+(n-1)\beta_2\sum\aix\ai=n$$
                                        $$\ssi\sum\aix\ai-\sum\ai\aix=n$$
                                        $$\ssi\sum[\ai,\aix]=-n$$
                                        But $\beta_2\not=0$, (or else 
                                        $1+\beta_1-\beta_2=0$),
                                        then from $(S)$ we get :
                                        $$-n\a1\ldots\an=\sum[\ai,\aix]\a1\ldots\an$$
                                        $$-n\a1\ldots\an=\sum_i([\ai,\aix]a_i a_1\ldots a_{i-1}a_{i+1}\ldots a_n)=0$$
                                        $$\Rightarrow \a1\ldots\an=0$$
                                        And we can iterate to get 
                                        $B=0$.
                                        \end{itemize}
                                \end{itemize}

        \item{$E_1\cup C\cup D$} : From $C\cup D$, the only remaining case to study is $\lambda=\mu$, that is to say 
        $n\nu=(\alpha+\beta)\lambda$. Moreover (by the case $E_2$), $\beta_2=0$, we must have $r+s=0$. Then we see that we can use $\xi$
        as in the case $E_2$, setting $\xi(\ai)=\xi(\aix)=x_i$, with
        $x_i^2={\displaystyle {\lambda\over n}}$.

        \item{$F$} : We have :
$$\matrix{
(r+s)\aj&=&r\aj(\alpha\sum\ai\aix+\beta\sum\aix\ai)+s(\alpha\sum\ai\aix+\beta\sum\aix\ai)\aj\cr
        &=&\alpha\sum_{i\not=j}\ai(r\aj\aix+s\aix\aj)+\beta\sum_{i\not=j}(r\aj\aix+s\aix\aj)\ai\hfill\cr
        & &\hfill+(r\beta+s\alpha)\aj\ajx\aj+r\alpha\aj^2\ajx+s\beta\ajx\aj^2\cr
        &=&(r\beta+s\alpha)\aj\ajx\aj+r\alpha\aj^2\ajx+s\beta\ajx\aj^2\cr}$$
      Since there always exists $j$ such that the monomials in the last expression are irreducible, we must have $r\alpha=s\beta=0$ and $r\beta+s\alpha=0$, that entails that either $r$ and $s$ or $\alpha$ and $\beta$ must be zero, which is impossible.
\end{itemize}
\smallbreak

\underline{$(1,1)_a\cup(1,-1)_a$} : This case is quite similar to the previous one, and we leave it to the reader. In the case $E_2$ we find the pseudo-boson solution.
\smallbreak

\underline{$(1,1)_c\cup (2,0)$ :}

\begin{itemize}
        \item{$A_3$} : Let us distinguish between the different forms of $S^{(0,0)}$.
                \begin{itemize}
                  \item{(a)} On one hand we have $\aj^2\ajx\rightarrow 0$, and on the other hand we find $\aj^2\ajx\rightarrow (1-\beta)\aj-\beta\sum_{i\not=j}\aj\aix\ai$, which is irreducible. This system cannot be confluent.
                   \item{(b)} : We can use the same argument by reducing $\ajz^2\ajzx$ in two different ways.
                \end{itemize}

        \item{$A_1\cup B_1\cup F$} : From $(l_3)\ai$ and $\ai(l_3)$ we get $\alpha=\beta$. We set $\alpha$ to $1$, and we define $\delta : L_n\rightarrow L_1=K\langle a,a^+\rangle $ by $\delta(\ai)=a$,
        $\delta(\aix)={\displaystyle a^+}$. This $\delta$ induces a non-zero morphism from $B$ to the Clifford algebra Cl$(1,1):=K\langle a,a^+\rangle/\langle a^2,{a^+}^2,aa^++a^+a-1\rangle$. Indeed, 
        $\delta(\ai\aix-\aj\ajx)=0=\delta(\aix\ai-\ajx\aj)$,
        $\delta(\ai\aj)=a^2=0$, $\delta(\aix\ajx)={\displaystyle
        {a^+}^2}=0$,
        $\delta(\ai\aix+\aix\ai)=aa^++a^+a=1$. We conclude by lemma \ref{lem7}.
        
        \item{$E_2$} :
$(l_1)\A2\Ax2\ssi\a1\ax1\A2\Ax2=\lambda\A2\Ax2$. Moreover, $\a1\ax1(l_2)\ssi
\a1\ax1\A2\Ax2=\lambda\a1\ax1$. Thus $\ai\aix=\aj\ajx$, which is not possible.

        \item{$E_1\cup C\cup D$} : From $C$ and $D$ we have $\lambda=\mu=0$. Then $n\nu=\alpha\lambda+\beta\mu=0$, which is impossible.

        \item{$F$} : We have $\alpha\sum_i\ai\aix\aj-\aj=0$ and $\beta\sum_i\aj\aix\ai-\aj=0$, but at least one one of these two expressions is irreducible.
\end{itemize}
\smallbreak

\underline{$(1,1)_b\cup(2,0)$} : 
        \begin{itemize}
        \item{$C$} : $B=0$ (see case $(1,1)_b\cup (1,-1)_a\cup (2,0)$).

        \item{$E_2$} : 
          \begin{itemize}
            \item{(a)} : We have $(\ai^2)\aix\r 0$, and on the other hand $\ai(\ai\aix)\r\lambda(1-\beta_1)\ai+\beta_1\beta_2{\sum_{\cal
                I}}\ajx\aj\ai+\beta_1\beta_2{\sum_{\cal
                J}}\ajx\ai\aj-\beta_2{\sum_{j\not=i}}\ai\ajx\aj$ where ${\cal I}$ is the set of indices $j$ s.t. $\aj<\ai$ and
                ${\cal J}$ the set of all $j$'s s.t. $\aj>\ai$. The last expression being irreducible, we come to a contradiction.
            \item{(b)} : We do the same with $\ai\aix^2$.
            \item{(c)} : ($n=2$) We have $\ajx\aj^2\r 0$, and also 
              $(\ajx\aj)\aj\r{{1\over\beta_2}}(\lambda\aj-\beta_1\aix\ai\aj-\ai\aix\aj)$. The term $\aix\ai\aj$ may be reduced to $\aix\aj\ai$, but the whole expression cannot reduce to $0$.
            \item{(d)} : We have $\aix\ai^2\r 0$ and $\aix\ai^2\r{1\over\beta_1}(\lambda(1-{1\over\beta_1})\ai-\beta_2\ajx\aj\ai+{\beta_2\over\beta_1}\ai\ajx\aj)$. Confluence implies that 
                $\beta_1=1$, $\beta_2=0$. In this case, $I$ is generated by the relations $\a1\ax1+\ax1\a1=\lambda$, $\A2\Ax2+\Ax2\A2=\lambda$, $\a1^2=\A2^2=\ax1^2=\Ax2^2=0$, $\ai\aj-\aj\ai=\aix\ajx-\ajx\aix=0$, $\forall i\not=j$. We can thus send $B$ onto $\Cl(1,1)$ by $\ai\mapsto a$,
                $\aix\mapsto a^+$. We then use lemma \ref{lem7}.
          \end{itemize}

          \item{$E_2'$} : \begin{itemize}
                            \item{(a)} : Same method as in (a) of 
                              case $E_2$.
                            \item{(b) or (c)} : We have $0\l\aiz^2\aizx\r\lambda\aiz-\sum_{\cal
                                I}\aiz\ai\aix-\sum_{\cal J}
                                \ai\aiz\aix$ 
                                with ${\cal I}=\{i|\aiz<\ai\}$ and
                                ${\cal J}=\{i|\aiz>\ai\}$. This contradicts confluence.
                          \end{itemize}

        \item{$F$} :
          \begin{itemize}
            \item{(a)} : We can assume without loss of generality that $i_0=n$, so that $\an\ai\r\ai\an$. One can then show that $0\l\an^2\anx\r{\lambda\over\alpha}(1-{\beta\over\alpha})\an+$ ${\beta\over\alpha}\sum_{i<n}\ai\aix\an+{\beta^2\over\alpha^2}\sum_{i<n}\aix\ai\an-{\sum_{i<n}\ai\an\aix-{\beta\over\alpha}\sum_{i<n}\an\aix\ai}$, which is irreducible.
           \item{(b)} : Symmetrical computation with $\anx^2\an$.
          \end{itemize}

          \item{$A_3$ or $A_1\cup B_1\cup F$} : It is easy to show that $\a1\ldots\an=0$. In any case the normal
              form of $\a1\ldots\an$ looks like 
              $a_{i_1}^{}\ldots a_{i_n}^{}\not=0$. Therefore, $(P_4)$ does not hold.
        \end{itemize}
\smallbreak

\underline{$(1,1)_a\cup(2,0)$} : This case is similar to the previous one.
\smallbreak

\underline{$(1,-1)_b\cup(2,0)$} :
         \begin{itemize}
        \item{$A_3$} : By multiplying $\a1\ax1+\beta\sum_i\aix\ai=1$ to the left by $\a1$, we find $\beta\A2\Ax2\a1=\a1$, 
                thus $B=0$. We use a similar method in cases $A_1\cup D$, $A_2\cup D$, and $A_1\cup B_1\cup F$.

        \item{$C$ or $C\cup D$} : We can assume that $i_0=n$. We have $0\l\an^2\anx\r\lambda\an-\an\a1\ax1-\ldots-\an a_{n-1}^{}a_{n-1}^+$, consequently the reduction system cannot be confluent.
        
        \item{$E_2$} : 
          \begin{itemize}
            \item{$n\geq 3$} : From $\a1(l_1)\A2$ we find 
              $\a1\A2=\beta_1\a1\ax1\a1\A2$, and from $\a1(l_3)\A2$, 
              $\a1\A2=\beta_2\a1\ax1\a1\A2$, thus $\a1\A2=0$.
            \item{$n=2$} : In every case we can easily prove that the reduction system is not confluent.
          \end{itemize}

        \item{$E_2'$} : Same methods as above.

        \item{$F$} : Let's assume we have a reduction system of type (a) (the case (b) is symmetrical), and suppose $i_0=n$. We have $0\l\an\anx\a1\r-\a1\ax1\a1-{\beta\over\alpha}\sum_{i>1}\aix\ai\a1+{1\over\alpha}\a1$, which is irreducible.
        \item{$E_1\cup C\cup D$} : $\forall i\in {\cal L}$,
          $0\l\ai\aix\aiz\r{{1\over\alpha}}(\nu\aiz-\beta\aix\ai\aiz)$, which is irreducible, and $\forall i\in {\cal K}$, $0\l\aiz\ajx\aj\r{{1\over\beta}}(\nu\aiz-\alpha\aiz\aj\ajx)$, irreducible too. Since at least one of two sets of indices is not empty, we conclude that the reduction system is not confluent.
        \end{itemize}
\smallbreak

\underline{$(1,-1)_a\cup(2,0)$} : 

        \begin{itemize}
        \item{$A_2\cup D$, $\lambda=0$} : If $\akx=\sup\{\aix\}$, we have
          $0\l\akx\ak^2\r\mu\ak-\sum_{i\not=k}\aix\ai\ak$, which is irreducible.

        \item{$C$ or $C\cup D$} : see $(1,-1)_b\cup (2,0)\cup C$.

        \item{$A_3$} : We know from proposition \ref{prop8} that at most one of the
        $\aix\ai$'s is reducible. But we have $0\l\ai^2\aix\r(1-\beta)\ai+\beta^2\sum_{j\not=i}\ajx\aj\ai-\beta\sum_{j\not=i}\ai\ajx\aj$. Now we can choose $i$ such that $\forall j\not=i$, $\ajx\aj$ is irreducible. Then if $r=0$, the last expression is irreducible. If
        $r\not=0$, some of the $\ai\ajx\aj$'s can be reduced to $q\ajx\ai\aj$, but in any case we get a non-zero irreducible quantity.

        \item{$A_1\cup B_1\cup F$} : $\ai(l_{3,1})\aj$ gives 
        $\ai(\alpha\a1\ax1+\beta\ax1\a1)\aj=\alpha\ai(\ai\aix)\aj+\beta\ai(\ajx\aj)\aj$, but this is excluded.

        \item{$A_1\cup D$} : We have, $\forall i$, $\ai\a1\ax1=\ai\ai\aix=0\Rightarrow \mu\a1\ax1=\sum\aix\ai\a1\ax1=0$. A contradiction.

        \item{$E_2$} : 
          \begin{itemize}
            \item{(a)} :
          By reducing $\a1^2\ax1$ in two different ways, one can prove that we must have $r\not=0$, $\beta_2=0$ and $\beta_1=1$  for the system to be confluent. Thus we get $\{\ai^2\r 0,\aix^2\r 0,\ai\ajx\r q\ajx\ai,\ai\aix\r 1-\aix\ai|1\leq i\not=j\leq n\}$ which is a 
        confluent reduction system adapted to $<_n$. Let $\II=(i_1,\ldots,i_k)\in[1..n]^k$ be a k-uple of indices. $\aii$ will stand for $a^{}_{i_1}\ldots a^{}_{i_k}$, and $\aiix$ for $a^+_{i_1}\ldots a^+_{i_k}$, $|\II|=k$. If $\II=\emptyset$, we set $a^{}_\emptyset=a^+_\emptyset=1$. With these notations, the basis of irreducible monomials for our reduction system is the set of all $\aiix\ajj$, $\II$ and $\JJ$ running over all possible t-uple of indices such that $i_m\not=i_{m+1}$, $j_m\not=j_{m+1}$, $\forall
        m$. It is then possible to use this basis to explicitly calculate the commutator of an element of $B^0$ with an $\ai$ and an $\aix$ (see \cite{bes1}). Doing this, one sees that $(P_3)$ does not hold.

       \item{(b)} : This case is symmetrical to the latter.
       \item{(c)} : $0\l\ajx\aj^2\r{1\over\beta_2}(\lambda\aj-\beta_1\aix\ai\aj-q'\ai\aj\aix)$, thus the reduction system is not confluent.
       \item{(d)} : $\aix\ai^2\r{1\over\beta_1}(\lambda(1-{1\over\beta_1})\ai-\beta_2\ajx\aj\ai+{\beta_2\over\beta_1}\ai\ajx\aj)$. Now this last expression is irreducible, except for the term $\ai\ajx\aj$, which can possibly be reduced to $q\ajx\ai\aj$. So, we must have $\beta_1=1$ and $\beta_2=0$ and then $rs\not=0$. Indeed, if $r=0$ we have $0\l\aj\ajx\ai\r\lambda\ai-\ajx\aj\ai$, and if $s=0$ we have $0\l\aj\aix\ai\r\lambda\aj-\aj\ai\aix$ both being impossible. There are two cases :
         \begin{itemize}
           \item{} If $\ai\ajx$ is reducible, then 
             $q\aix\ajx\ai\l\aix\ai\ajx\r\lambda\ajx-\ai\aix\ajx$. Since both expressions are irreducible, we conclude that the system is not confluent.
            \item{} If $\ajx\ai$ is reducible, then we have 
              $q'\aj\ai\ajx\l\aj\ajx\ai\r\lambda\ai-\ajx\aj\ai$, and arrive at the same conclusion.
         \end{itemize}
\end{itemize}

        \item{$E_2'$} : Case (b) or (c) $0\l\aiz^2\aizx\r\lambda\aiz-{\displaystyle\sum_{j\not=i_0}}\aiz\aj\ajx$, which is irreducible.

        \item{$E_1\cup C\cup D$} :
          $\lambda'\aiz+{\beta\over\alpha}\sum_{\cal
            L}\akx\ak\aiz-\sum_{\cal K}\ak\akx\aiz\l\aiz\aizx\aiz\r\mu'\aiz+{\alpha\over\beta}\sum_{\cal
           K}\aiz\ak\akx-\sum_{\cal L}\aiz\akx\ak$ and the only further reductions we can possibly do are : $\ak\akx\aiz\rightarrow
           q'\ak\aiz\akx$ and $\aiz\akx\ak\rightarrow q\akx\aiz\ak$. This shows that the two expressions cannot be reduced to a common normal form.
        
        \item{$F$} : Let's do the case $(a)$, the other one being symmetrical. Suppose $\an$ is the largest of the $\ai$'s, then $0\l\an^2\anx\r\an(1-\alpha\sum_{i<n}\ai\aix-\beta\sum_{i\leq n}\aix\ai)$. If $r=0$ we see that confluence is impossible. And if 
        $r\not=0$, we have $0=\an-\alpha\sum_{i<n}\an\ai\aix-q\beta\sum_{i<n}\aix\ai\an$ and we come to the same conclusion.
        \end{itemize}
\smallbreak

\underline{$(1,-1)_b$} :
        \begin{itemize}
        \item{$A_3$} : From $\a1\an=\a1(\alpha\a1\ax1+\beta\sum_i\aix\ai)\an$ we are led to $\a1\an=0$. Thus, the presentation is not standard. The same method can be used for the cases $A_1\cup D$, $A_2\cup D$, $\lambda=0$ and $A_1\cup B_1\cup F$.

        \item{$C$, $C\cup D$, $E_2$, $E_2'$, $E_1\cup C\cup D$, $F$} : These rather easy cases are left to the reader (it can be shown each time that the systems cannot be confluent).
        \end{itemize}
\smallbreak

\underline{$(1,-1)_a$} : Let us first notice that in all cases containing $A_2$ we have $r\ai\ajx+s\ajx\ai=0\Rightarrow r\aj\ai\ajx+s\lambda\ai=0$. Then, if $r=0$, $\ai=0$, and
                        $(P_0)$ is not fulfilled. So  $r\not=0$, and $\ai\ajx$ must be
                          reducible.
\begin{itemize}
        \item{$A_2$} : According the remark, we must have $r\not=0$. Then we see that
                        $S=\{\ai\aix\r 1,\ai\ajx\r q\ajx\ai|i\not=j\}$ is confluent (there are no ambiguities) and adapted to $<_n$. The basis of irreducible monomials is 
                        $T:=\{\aiix\ajj|\II,\JJ$ run over all tuples of indices$\}$. If we write $p$ for the projection on
                        $W=$Span$(T)$ in the direction of $I$, then 
                        $p(x)=$nf$(x)$. If the lenghts of $\II$ and  
                        $\JJ$ are at least 2, and if $n$ belongs to
                        $\JJ$, we have 
                        $p([\aiix\ajj,\anx])=q^{n_\JJ}\aiix
                        a^{}_{\JJ'}-\anx\aiix\ajj$, where $n_\JJ$ is the number of $\ai$'s in  
                        $\ajj$ that are to the right of the $\an$ which is the most to the right, and $\JJ'=\JJ$ where the $n$ the most to the right has been left out. If $n$ does not appear in $\JJ$,
                        $p([\aiix\ajj,\anx])=q^{|\JJ|}\aiix\anx\ajj-\anx\aiix\ajj$. In both cases, if $\II\not=(n,\ldots,n)$, we have lm$(p([\aiix\ajj,\anx]))=\anx\aiix\ajj$. Now if $\aiix\ajj\in L_n^0$, $\II=(n,\ldots,n)\Rightarrow \JJ=(n,\ldots,n)$, then if $\aiix\ajj\in L_n^0$ we always have lm$(p([\aiix\ajj,\anx]))=\anx\aiix\ajj$. Let $\tilde{N_1}$ be a representative of $N_1$ in $W$, and let $\aiix\ajj$ be the leading term of $\tilde{N_1}$. Then we see that $\anx\aiix\ajj$ is bigger than any other term of $p([\tilde{N_1},\anx])$, thus lm$(p([\tilde{N_1},\anx]))=\anx\aiix\ajj$, which proves that $(P_3)$ does not hold.
                    
        \item{$A_2\cup B_1$, $A_2\cup B_2$} : As before, we have 
        $r\not=0$. For $i\not=i_0$, we get 
        $\ai=(\ai\aix)\ai=\ai(\aizx\aiz)=q\aizx\ai\aiz$. Then
        $\ai-q\aizx\ai\aiz=0$, which is irreducible.

        \item{$A_2\cup D$} : It can readily be seen that we need $\lambda=\mu$ and $q=0$ in order to have a confluent system. In this case, an argument like the one we used in case $A_2$ allows us to prove that $(P_3)$ is not satisfied.

        \item{$A_3$} : 
           \begin{itemize}
               \item{If $r\not=0$} :
                $S=\{\ai\aix\r 1-\beta\sum\ajx\aj,\ai\ajx\r q\ajx\ai|i\not=j\}$ is confluent and adapted to $<_n$ (no ambiguities). Using the basis $T=\{\aiix\ajj\}$ it can be shown that $(P_3)$ does not hold (see \cite{bes1}).

            \item{si $r=0$} : Let's look at the two possible reduction systems :
                \begin{itemize} 
                \item{(a)} :  we have $0\l\an\anx\a1\r\a1-\beta\sum_i\aix\ai\a1$, and there is no confluence.
                \item{(b)} : For $i\not=i_0$,
                  $0\l\ai\aix\aiz\r\aiz\aizx\aiz$, which is irreducible.
                \end{itemize}
           \end{itemize}

        \item{$A_1\cup B_1\cup F$} : 
          \begin{itemize}
            \item{(a)} We have $\aiz\aizx\ai\leftarrow \ai\aix\ai\rightarrow{\displaystyle {1\over\beta}}(\ai-\alpha\ai\aiz\aizx)$. The term on the left may be reducible to 
          $q'\aiz\ai\aizx$. In any case, the ambiguity is not solvable.
            \item{(b)} : With $i\not=i_0$ we have ${1\over\alpha}(\ai-\beta\aizx\aiz\ai)\leftarrow\ai\aix\ai\rightarrow\ai\aizx\aiz$. We conclude as before.
          \end{itemize}
        \item{$A_1\cup D$} : If $\ai\ajzx$ is reducible then $q\ajzx\ai\ajz\leftarrow\ai\ajzx\ajz\rightarrow\ai-{\sum_{j\not=j_0}}\ai\ajx\aj$. Now these two expressions cannot be reduced to a common form, so $\forall i\not=j_0$, $\ai\ajzx$ is 
            irreducible. Consequently $\ajzx\ai$ is reducible.
            \begin{itemize}
              \item{}If $n\geq 3$, or $n=2$ and $i_0=j_0$, we can take 
                $i\not=i_0,j_0$. We have $q'\ai\ajzx\aix\leftarrow\ajzx\ai\aix\rightarrow\ajzx\aiz\aizx$. Then if $i_0\not=j_0$, $\ajzx\aiz\aizx\rightarrow q'\aiz\ajzx\aizx$ and the reduction system is not confluent. If $i_0=j_0$, $\aizx\aiz\aizx\rightarrow\aizx-{\sum_{j\not=i_0}}\ajx\aj\aizx$, which is irreducible.\newline
            We leave the case $n=2$ and $i_0\not=j_0$ to the reader (the system is not confluent).
            \end{itemize}

        \item{$C$} : As in case $(1,1)_c\cup(1,-1)_b$ the homomorphism $\phi$ can be defined and used to rule out this case.

        \item{$C\cup D$} :
          \begin{itemize}
            \item{}If $\exists k\not=j_0$ such that $\ak\ajzx$ is reducible, then
               $q\ajzx\ak\ajz\l\ak\ajzx\ajz\r$ $\mu\ak-q{\sum_{j\in{\cal K}}}\ajx\ak\aj-{\sum_{j\in{\cal L}}}\ak\ajx\aj-\ak\akx\ak$, with ${\cal K}=\{j\not=j_0|\ak\ajx$ reducible$\}$, and
                ${\cal L}=\{j\not=j_0|\ak\ajx$irreducible$\}$. We see that the expression is irreducible if $k\not=i_0$. If $k=i_0$, it can be reduced to
$(\mu-\lambda)\ak-q\sum_{j\in{\cal K}}\ajx\ak\aj-\sum_{j\in{\cal L}}\ak\ajx\aj+\!\sum_{j\in{\cal K'}}\aj\ajx\ak+q'\sum_{j\in{\cal L'}}\aj\ak\ajx$. Therefore we must have $\mu=\lambda$. But in this case we can use $\phi$, as in case $(C)$, and $(P_3)$ is not satisfied. Thus we must have $\forall k\not=j_0$, $\ak\ajzx$ is irreducible, which entails $\ajzx\ak$ is reducible.
            \item{} If $\exists k\not=i_0$ such that $\aizx\ak$ is reducible, we can reduce $\aiz\aizx\ak$ in two different ways, and come to a contradiction.
              \item{} If $n>2$, or $n=2$ and $i_0=j_0$.\newline
Let 
                $k\not=i_0,j_0$, then $q'\ak\ajzx\aizx\l\ajzx\ak\aizx\r q\ajzx\aizx\ak$. Since these two expressions are irreducible, this case is ruled out.
              \item{} $n=2$, $i_0=1$, $j_0=2$. We leave this case to the reader (reduce $\Ax2\A2\ax1$ in two different ways).
          \end{itemize}

        \item{$E_2$ or $E_2'$} 
          \begin{itemize}
            \item{}If $r\not=0$,
        $S=\{\ai\aix\r 1-\beta_1\aix\ai-\beta_2\sum\ajx\aj,\ai\ajx\r q\ajx\ai|i\not=j\}$ is confluent and adapted to $<_n$. $T=\{\aiix\ajj\}$ is the corresponding basis of irreducible monomials. By using the same method as in case $A_3$, we find that $(P_3)$ does not hold.
            \item{}If $r=0$,
              $S=\{\aix\aj\r 0,\aix\ai\r 1-\alpha_1\ai\aix-\alpha_2\sum_{k\not=j}\ak\akx\}$ is confluent and adapted to $<_n$. The corresponding basis is $T=\{\aii\ajjx\}$. If ${\cal I}$ does not begin with $n$, we have $[\aii\ajjx,\anx]=\aii\ajjx\anx$. Moreover, $[\an\aii\ajjx,\anx]=\an\aii\ajjx\anx-\aii\ajjx+\alpha_1\an\anx\aii\ajjx+\alpha_2\sum_{i<n}\ai\aix\aii\ajjx$. Thus we see that if ${\cal I}$ does not begin with $n$, lm$([\an\aii\ajjx,\anx])=\an\aii\ajjx\anx$, and we get by induction : if ${\cal I}\not=(n,\ldots,n)$, lm$([\an\aii\ajjx,$ $\anx])=\an\aii\ajjx\anx$, and $[\an^k\anx^k,\anx]=(1-(-1)^k\alpha_1^k)\an^k\anx^{k+1}+u$, with $u<\an^k\anx^{k+1}$. Then if lm$(N_1)\not=\an^k\anx^k$,  lm$([N_1,\anx])=$lm$(N_1)\anx$, which is irreducible : a contradiction. And if lm$(N_1)=\an^k\anx^k$, we show that lm$([N_1,\ax1^k])=\an^k\anx^k\ax1^k$, which is irreducible as well. Thus $(P_3)$ cannot hold.
          \end{itemize}

        \item{$E_1\cup C\cup D$} : 
                \begin{itemize}
                \item{$r=0$} : If $i\in {\cal L}$,
                  $0=\ai\aix\aj={{1\over\alpha}}(\nu\aj-\beta\aix\ai\aj)$, contradiction. Thus ${\cal L}=\emptyset$. But now, with $j\not=i_0$, $\aiz\aizx\aj=\lambda\aj-{\sum_{j\not=i_0}}\ai\aix\aj=\lambda\aj-\aj\ajx\aj$, which also contradicts the confluence.
                \item{$s=0$} : Symmetrical case.
                \item{$rs\not=0$} : We have $\aiz>\ajx$, $\forall
                  j\in{\cal L}$, and $\aizx>\aj$, $\forall j\in{\cal
                  K}$. The reader can then easily verify that $\aizx\aiz\ajx$ can be reduced in two different ways which lead to distinct irreducible expressions.
                \end{itemize}

        \item{$F$} :
                \begin{itemize}
                \item{$\alpha\not=-\beta$} : We define an homomorphism $\phi : \bigoplus K[x_i,y_i]/\langle\{x_iy_i-{1\over\alpha+\beta}\}\rangle$, and we use lemma \ref{lem4}.
                \item{$\alpha=-\beta$} : Suppose $\alpha=1$. We have 
                $\sum\ai\aix-\sum\aix\ai=1$\newline
                If $r=0$ then $\ai\aix\ai-\sum_j\ajx\aj\ai=\ai$, and
                $\ai\sum_j\aj\ajx-\ai\aix\ai=\ai$. But both expressions are irreducible. This is impossible.\newline
                Now if $r\not=0$,
                $S=\{\an\anx\r 1+\sum\aix\ai-\sum_{i<n}\ai\aix,\ai\ajx\r q\ajx\ai|i\not=j\}$ is confluent and adapted to $<_n$. The basis $T$ is made of monomials of the form $a^+_{\II_1}a^{}_{\JJ_1}a^{}_{i_1}a^+_{i_1}a^+_{\II_2}a^{}_{\JJ_2}a^{}_{i_2}a^+_{i_2}\ldots$, with $i_1,i_2,\ldots\not=n$. If $t$ is such a monomial we have $[t,\anx]=a^+_{\II_1}a^{}_{\JJ_1}\ldots(a^{}_{i_j}a^+_{i_j})\ldots a^+_{\II_k}a^{}_{\JJ_k}\anx-\anx a^+_{\II_1}a^{}_{\JJ_1}\ldots$ Set $t=$lm$(\tilde N_1)$.
	\begin{itemize}
		\item{}If $t=\ajx u$, $j\not=n$, then lm$(p([t,\aj]))=\aj t$, thus $(P_3)$ does not hold.
		\item{}If $t=\anx u$, then lm$(p([t,\an]))=a^{}_{n-1}a^+_{n-1} u$, $(P_3)$ does not hold.		 
		\item{}If $t=\aj u$, lm$(p([t,\an]))=\an t$, $(P_3)$ does not hold.
	\end{itemize}
	\end{itemize}
       \end{itemize}
                
\underline{$(1,1)_c$} : This case is rather easy, and we only state the results.
        \begin{itemize}
        \item{$A_3$, $A_3\cup B_1$, $A_1\cup D$, and $A_2\cup D$} : It is easily shown that $\ai^3=0$, but since $\ai^3$ is irreducible, this contradicts $(P_4)$.

        \item{$C$} : Use the same $\phi$ as in case 
        $(1,1)_c\cup(1,-1)_b$.

        \item{$C\cup D$} : If $\lambda=\mu$, we can do the same as above. If $\lambda\not=\mu$, it can be shown that $\ai^3=0$.

        \item{$E_2$, $E_2'$, $F$, $E_1\cup C\cup D$} : $(P_4)$ does not hold. 

        \end{itemize}
\smallbreak

\underline{$(1,1)_b$} :
        \begin{itemize}
        \item{$A_2$, $A_2\cup B_1$, $A_2\cup B_2$} : Let $\phi : B\rightarrow
        C$, $C:=K[x,y]/\langle xy-1\rangle $, be the only homomorphism such that $\phi(\ai)=x$, $\phi(\aix)=y$. $\phi$ is clearly well defined and we conclude by lemma \ref{lem4}.

        \item{$A_2\cup D$} : With $i_0=n$, we have $\lambda\an\leftarrow\an\anx\an\rightarrow\mu\an+\sum_{i<n}\an\aix\ai$. Thus $(P_4)$ cannot hold.

        \item{$A_3$} :
                \begin{itemize}
                \item{$\beta\not=-1/n$} : As in $A_2$ we can define $\phi :
                B \rightarrow C:=K[x,y]/\langle xy-1/(1+n\beta)\rangle $.
                \item{$\beta=-1/n$} : We define $\phi :
                B\rightarrow A_1=K\langle a,a^+\rangle /\langle aa^+-a^+a-1\rangle $ by setting 
                $\phi(\ai)=a$, $\phi(\aix)=a^+$. We then use lemma \ref{lem7}.
                \end{itemize}

        \item{$A_1\cup B_1\cup F$} : Same method as above.

        \item{$A_1\cup D$} : We send $B$ to 
        $K[x,y]/\langle xy-1/n\rangle $ by $\ai\mapsto x$, $\aix\mapsto y$, and we use lemma \ref{lem4}.

        \item{$C$} : Here we send $B$ to ${ \bigoplus_i}
        K[x_i,y_i]/\langle x_iy_i-1\rangle $, as in case $(1,1)_c\cup (1,-1)_b$.

        \item{$C\cup D$} : We are going to show that $(P_3)$ does not hold if $n=2$. We will then be able to deduce that $(P_3)$ never holds with the help of lemma \ref{lem5}, since the projection $L_n\rightarrow L_2$ goes over to the 
          quotient.\newline
          $S=\{\a1\ax1\r\lambda-\A2\Ax2,\Ax2\A2\r\mu-\ax1\a1,\a1\A2\r\A2\a1,\Ax2\ax1\r\ax1\Ax2\}$ is confluent (no ambiguity) and adapted to the deglex ordering coming from $\A2<\a1<\ax1<\Ax2$. The corresponding basis is $T=\{\A2^{k_1}\ax1^{l_1}\ldots\A2^{k_r}\ax1^{l_r}\a1^{m_1}\Ax2^{n_1}\ldots\a1^{m_s}\Ax2^{n_s}\}$, with $k_i,l_i,m_i,n_i\geq 0$.\newline
Let $x\in T\cap L_2^{(0,0)}$, $x\not=1$. Then $x$ begins with an $\A2$ or an $\ax1$, and ends with an $\a1$ or an $\Ax2$, therefore $x\ax1$ is reducible, meanwhile $\ax1 x$ is irreducible. Moreover $\ax1 x>x\ax1$, so lm$(p([N_1,\ax1]))=\ax1$lm$(N_1)$. This is absurd : $(P_3)$ cannot hold.

        \item{$E_2$ or $E_2'$} : This case looks like $A_3$. We have to use different kinds of morphisms depending on whether $\beta_1+(n-1)\beta_2+1$ vanishes or not. We leave the details to the reader.

        \item{$E_1\cup C\cup D$} : Let's deal with the case ${\cal
            K}\not=\emptyset$, the case ${\cal L}\not=\emptyset$ being symmetrical.
            Let $i$ be such that $\ai=$min$\{\aj|1\leq
            j\leq n\}$. We must have $i\not=i_0$. We get $\aizx\ai\aiz\l\aizx\aiz\ai\r\mu'\ai-\sum_{\cal
            L}\ajx\ai\aj+{{\alpha\over\beta}}\sum_{\cal
            K}\aj\ajx\ai$ and this contradicts $(P_4)$.

        \item{$F$} : Quotienting out by the ideal generated by 
          $\sum\ai\aix-\lambda_1$ and $\sum\aix\ai-\lambda_2$, with
          $\lambda_1+\lambda_2=\lambda\not=0$, we are brought back to case
          $C\cup D$, and we conclude by lemma \ref{lem5}.
        \end{itemize}
\smallbreak

\underline{$(1,1)_a$} :
        \begin{itemize}
        \item{$A_2$, $A_2\cup B_1$, $A_2\cup B_2$} : Let $i,j$ be s.t. $\ai>\aj$. Then $-\aj\ai\ajx\leftarrow\ai\aj\ajx\rightarrow\ai$ and $(P_4)$ cannot hold.

        \item{$A_2\cup D$} : We have $\lambda\aiz\leftarrow\aiz\aizx\aiz\rightarrow\mu\aiz-\sum_{i\not=i_0}\aiz\aix\ai$ and $(P_4)$ cannot hold.

        \item{$A_1\cup D$} : We define $\delta : B\rightarrow
          \Cl(n,0):=K\langle x_1,\ldots,x_n\rangle/\langle x_ix_j+x_jx_i-\delta_{ij}\rangle$ by
          $\delta(\ai)=\delta(\aix)={{1\over\sqrt
          n}}x_i$, we then use lemma \ref{lem6}.

        \item{$A_3$} :
          \begin{itemize}
            \item{(a)} : Suppose that $\ai$ is the sup of the $\ak$'s, and choose $j\not=i$. We find : $-\aj\ai\ajx\leftarrow\ai\aj\ajx\rightarrow{ \!(1\!-\!\beta)\ai+\beta^2\!\sum\!\akx\ak\ai\!-\!\beta\!\sum_{k\not=i}\!\ai\akx\ak}$. This contradicts $(P_4)$.
            \item{(b)} : If $n\geq 3$, let $i,j\not=i_0$
              be s.t. $\ai<\aj$. Then $-\ai\aj\aix\l\aj\ai\aix\r$ $-\aiz\aj\aizx$, which contradicts $(P_4)$. If $n=2$, one can easily show that $i_0\not=j_0$ leads to a contradiction. We can then assume that $i_0=j_0=2$. If we define $<$ by $\A2<\a1<\ax1<\Ax2$, we see that $S=\{\a1\A2\r-\A2\a1,\Ax2\ax1\r-\ax1\Ax2,\a1\ax1\r\A2\Ax2,\Ax2\A2\r\lambda-\alpha\A2\Ax2-\ax1\a1\}$ is confluent and adapted to $<$.\newline
$T=\{\A2^{k_1}\ax1^{l_1}\ldots\A2^{k_r}\ax1^{l_r}\a1^{m_1}\Ax2^{n_1}\ldots\a1^{m_s}\Ax2^{n_s}\}$. We proceed exactly as in case $(1,1)_b\cup C\cup D$.

          \end{itemize}

        \item{$A_1\cup B_1\cup F$} : It can easily be shown that neither the system (a) nor the system (b) is confluent.

        \item{$C$ or $C\cup D$} : We do as in case $(1,1)_b$.

        \item{$E_2$ or $E_2'$} : Once again, if $n\geq 3$, we can easily show that $(P_4)$ does not hold. If $n=2$, we find a confluent reduction system, and we show that $(P_3)$ does not hold.

        \item{$E_1\cup C\cup D$} : See $(1,1)_b$ ($(P_4)$ does not hold).

        \item{$F$} : See $(1,1)_b$ ($(P_3)$ does not hold).
        \end{itemize}
\medbreak

This is the end of the demonstration for the case $n\geq 2$. Let us know look at the case $n=1$.

The only relations we can have are : $(2,0)$, $(A_2)$, $(B_2)$, and
$(F)$ (by lemma \ref{lem1}).

\underline{$(2,0)$} : 

        \begin{itemize}
        \item{$A_2$} : $a^2a^+=0=a\Rightarrow B=0$

        \item{$B_2$} : Symmetrical to the above case.

        \item{$F$} : Multiplying $\alpha aa^++\beta a^+a=1$ to the right, then to the left by $a$ we get $\alpha=\beta$, and we recongnize the Clifford algebra $C_1$ which is a solution to our problem.
        \end{itemize}

\underline{$A_2$ or $B_2$} : We can send $B$ to 
        $C=K[x,y]/\langle xy-1\rangle $, by a non-zero isomorphism, and we conclude by lemma \ref{lem4}.

\underline{$F$} : If $\alpha+\beta\not=0$, we can send $B$ to
$K[x,y]/\langle xy-(1/\alpha+\beta)\rangle $ and use lemma \ref{lem4}.\newline
If $\beta=-\alpha$, we find the Weyl algebra $A_1$, which is the bosonic solution.

\hfill QED.

\section{Number Operator Algebras of Infinite Type}
\subsection{The Classification Theorem}

\begin{theorem}\label{theo2}
Let $\alpha$ be an  infinite cardinal number and let $B=L_\alpha/I$ be a symmetric n.o.a. of quadratic presentation, i.e. $I$ satisfies the following properties 
\item{$(P_0)$} $I\not=L_\alpha$, $I\not=\langle X\rangle$.
\item{$(P_1)$} $J(I)\subset I$.
\item{$(P_2)$} $\forall\sigma\in{\cal S}_\alpha$, $\sigma^*(I)\subset I$.
\item{$(P_3)$} $\forall i\in\II_\alpha$, $\exists N_i\in B$ s.t. (\ref{eq1}) and (\ref{eq2}) hold.
\item{$(P_4')$} : $I$ is generated by elements of degree two or less.

Then there exists $h\in R\setminus\{0\}$ such that $I$ is generated by one of the following sets :
\item{(a)} $\{\ai^2,\aix^2,\ai\aj+\aj\ai,\aix\ajx+\ajx\aix,\ai\ajx+\ajx\ai,\ai\aix+\aix\ai-h| i\not=j\}$
\item{(a')} $\{\ai^2,\aix^2,\ai\aj-\aj\ai,\aix\ajx-\ajx\aix,\ai\ajx-\ajx\ai,\ai\aix+\aix\ai-h|i\not=j\}$
\item{(c)} $\{\ai\aj-\aj\ai,\aix\ajx-\ajx\aix,\ai\ajx-\ajx\ai,\ai\aix-\aix\ai-h|i\not=j\}$
\item{(c')} $\{\ai\aj+\aj\ai,\aix\ajx+\ajx\aix,\ai\ajx+\ajx\ai,\ai\aix-\aix\ai-h|i\not=j\}$
\end{theorem}

The algebras of case $(a)$ (resp. $(a')$, $(c)$, $(c')$) are called fermionic (resp. pseudo-fermionic, bosonic, pseudo-bosonic) algebras, and are denoted by $\hat C_\alpha$ (resp. $C_\alpha$, $A_\alpha$, $\hat A_\alpha$).

\subsection{The Lemmas}
The lemmas 1, 2, 4, 5, 6, 7, and 8  are valid both in the finite and infinite cases.

Due to the action of the symmetric group, the terms with a sum cannot survive in the infinite case, or else the sum should be infinite, which is meaningless in our purely algebraic setting. For this reason, the lemma 10 gets replaced by :

\begin{lemma}
If $I$ fulfils $(P_0)$, $(P_1)$, $(P_2)$, $(P_3)$ and $(P_4')$ then it is generated by a union of sets, each being of one of the forms (with $\alpha$, $\beta$, $\lambda$, $r$, $s\in R$) :
\begin{itemize}
\item{ $(2,0)$ :} $\{\ai^2,{\aix}^2|i\in\II_\alpha\}$
\item{ $(1,1)_a$ :}
  $\{\ai\aj+\aj\ai,\aix\ajx+\ajx\aix|i,j\in\II_\alpha,i\not=j\}$
\item{ $(1,1)_b$ :} $\{\ai\aj-\aj\ai,\aix\ajx-\ajx\aix|i,j\in\II_\alpha,i\not=j\}$
\item{ $(1,1)_c$ :} $\{\ai\aj,\aix\ajx|i,j\in\II_\alpha,i\not=j\}$
\item{ $(1,-1)_a$ :} $\{r\ai\ajx+s\ajx\ai|i,j\in\II_\alpha,i\not=j\}$,
  $(r,s)\not=(0,0)$, $r,s\in R$.
\item{ $(1,-1)_b$ :} $\{\ai\ajx,\ajx\ai|i,j\in\II_\alpha,i\not=j\}$
\item{ $A_1$ :} $\{\ai\aix-\aj\ajx|i,j\in\II_\alpha,i>j\}$
\item{ $A_2$ :} $\{\ai\aix-\lambda|i\in\II_\alpha\}$
\item{ $E$ :}
  $\{\alpha\ai\aix+\beta\aix\ai-\lambda|i\in\II_\alpha\}$
\item{ $E_1$ :} $\{\alpha(\ai\aix-\aj\ajx)+\beta(\aix\ai-\ajx\aj)|i,j\in\II_\alpha,i\not=j\}$
\end{itemize}
and the forms $B_1$, $B_2$ symmetric to $A_1$, $A_2$, by the exchange of $\ai$ and $\aix$.
\end{lemma}

We must combine these different sets of generators to enumerate all possible presentations. The equivalent of the proposition 1  is the following :

\begin{proposition}\label{prop4}
If $I$ fulfils $(P_0)$, $(P_1)$, $(P_2)$, $(P_3)$ and $(P_4')$ then there exists a presentation $R$ of
$I$, of the form $R=R^{(2,0)}\coprod R^{(1,1)}\coprod R^{(1,-1)}\coprod\penalty -100 R^{(0,0)}$, such that :
        \begin{itemize}
                \item{}$R^{(2,0)}=(2,0)$ or the empty set.
                \item{}$R^{(1,1)}=(1,1)_a$ or $(1,1)_b$ or $(1,1)_c$ or $\emptyset$.
                \item{}$R^{(1,-1)}=(1,-1)_a$ or $(1,-1)_b$ or $\emptyset$.
        \end{itemize}
\smallbreak
and $R^{(0,0)}$ is one of the following sets :

$A_2=\{\ai\aix-\lambda|i\in\II_\alpha\}$

$A_2\cup B_1=\{\ai\aix-\lambda,\ajx\aj-\azx\az|i,j\in\II_\alpha,j>0\}$

$A_2\cup B_2=\{\ai\aix-\lambda,\aix\ai-\lambda|i\in\II_\alpha\}$

$E=\{\ai\aix+\beta\aix\ai-\lambda|i\in\II_\alpha\}$

$A_1\cup B_1\cup E=\{\ai\aix-\az\azx,\aix\ai-\azx\az,\az\azx+\beta\azx\az-\lambda|i>0\}$

as well as $B_2$, $B_2\cup B_1$. In each case $\lambda$ and $\beta$ are non-zero, and belong to $R$.
\end{proposition}

\subsection{Sketch of Proof of Theorem 2}

We define a well-ordering $<$ on $X$ by $\ai<\aj\ssi\aix<\ajx\ssi i<j$ and $\aix<\aj$ for all $i,j\in\II_\alpha$. We call $<_n$ the deglex-ordering defined by $<$.
\medbreak

As we did in the finite case, we can easily get rid of the ideals that contain relations $(1,1)_c$, $(1,-1)_b$ or $(2,0)$, on the one hand, and $A_2$, on the other hand. Furthermore, the relations $(2,0)$ together with $E$ or $A_1\cup B_1\cup E$ imply $\beta=1$, as is easily seen by multiplying $\az\azx+\beta\azx\az=1$ on the left, then on the right, by $\az$.

\medbreak 

\underline{$(1,1)_c\cup(1,-1)_b\cup(2,0)$} : The only case to study is $E$. If we multiply the relation $E$ on the left by $\aj$, with $i\not=j$, we get $\lambda\aj=0$. Thus $B=0$, in contradiction with $(P_0)$.
         
\smallbreak
\underline{$(1,1)_{a\mbox{ or }b}\cup(1,-1)_b\cup(2,0)$} :
     \begin{itemize}
      \item{$E$ or $A_1\cup B_1\cup E$} : Let us multiply $\ak\akx+\beta\akx\ak=\lambda$ on the left by $\ai$ and on the right by $\aj$, with $i,j,k$ distinct. We get $\ai\aj=0$, and we are back to the previous case.
     \end{itemize}
\smallbreak

\underline{$(1,1)_c\cup (1,-1)_a\cup (2,0)$} :
     \begin{itemize}
       \item{$E$, $E\cup A_1\cup B_1$} : We have $rs=0$ (see the finite case $F$). Now if $s=0$, we see that $\lambda\ai=\ai\aj\ajx+\beta\ai\ajx\aj$ vanishes if $i\not=j$. The case $r=0$ is symmetrical.
       \end{itemize}
\smallbreak

\underline{$(1,1)_a\cup(1,-1)_a\cup (2,0)$} :
     \begin{itemize}
       \item{$E$}: We have (with $i,j,k$ distinct) :
$$\lambda r\ai\ajx=r(\ai\ak\akx\ajx+\beta\ai\akx\ak\ajx)=s\lambda\ai\ajx$$
$$\Rightarrow(r-s)\ai\ajx=0$$
If $\ai\ajx=0$, we easily find that $B=0$, thus we can assume $r-s=0$. We also have :
$$\left\{\matrix{\lambda\ai=\ai^2\aix+\beta\ai\aix\ai=\beta\ai\aix\ai\cr
\lambda\ai=\ai\aix\ai+\beta\aix\ai^2=\ai\aix\ai}\right.$$
So $\beta=1$ by $(P_0)$. Thus $B$ is a fermionic algebra $\hat C_\alpha$.
       \item{$E\cup A_1\cup B_1$}: Thanks to the previous case we have $r-s=0$ and $\beta=1$. We let the reader use the relations $A_1$ and $B_1$ to show that $B=0$.
     \end{itemize}
\smallbreak

\underline{$(1,1)_b\cup(1,-1)_a\cup(2,0)$} : This case is similar to the preceding one, except that we find pseudo-fermionic instead of fermionic algebras.
\smallbreak

\underline{$(1,1)_c\cup(1,-1)_b$} : We can do as in case $(1,1)_c\cup(1,-1)_b\cup(2,0)$.
\smallbreak

\underline{$(1,1)_{a\mbox{ or }b}\cup (1,-1)_b$} : We must have $\ai\aj=0$ (see the finite case).
\smallbreak

\underline{$(1,1)_c\cup(1,-1)_a$} : 
     \begin{itemize}
       \item{$E$} : We have : $r\ai\ajx=r(\ai\ak\akx\ajx+\beta\ai\akx\ak\ajx)=-s\beta\ai\akx\ajx\ak=0$. It entails that $s\ajx\ai=0$. If $\ai\ajx=0$, we get $\lambda\ai=\ai\aj\ajx+\beta\ai\ajx\aj=0$, thus $B=0$. This is the same if $\ajx\ai=0$.
     \end{itemize}
\smallbreak

\underline{$(1,1)_b\cup (1,-1)_a$} :
     \begin{itemize}
       \item{$A_2$, $A_2\cup B_1$, $A_2\cup B_2$} : If $r+s\not=0$, we find $B=0$ (see the finite case). If 
         $r+s=0$, we define $\delta : B\rightarrow
         C:=K[(x_i)_{i\in\II_\alpha},(y_i)_{i\in\II_\alpha}]/\langle x_iy_i-1\rangle$ by
         $\ai\mapsto x_i$ and $\aix\mapsto y_i$. Thus $(P_3)$ cannot hold, by lemma 5.
       \item{$E$} : According to the finite case, we must have $r+s=0$.
         \begin{itemize}
           \item{If $\beta+1=0$} : We find a bosonic algebra $A_\alpha$.
           \item{If $\beta+1\not=0$} : We can non-trivially map 
             $B$ onto $K[(x_i)_{i\in\II_\alpha}]/\penalty -100\langle x_i^2-1/(1+\beta)\rangle$ by 
             $\ai,\aix\mapsto x_i$. Thus $(P_3)$ cannot hold.
         \end{itemize}
       \item{$E\cup A_1\cup B_1$} : Of course we must also have $r+s=0$. If $\beta+1=0$, we have a quotient of a bosonic (Weyl) algebra, which is simple. Thus $B=0$. If $\beta+1\not=0$, one shows that $Z(B)=B^0$ (see the finite case). Thus $(P_3)$ cannot hold.
     \end{itemize}
\smallbreak

\underline{$(1,1)_a\cup(1,-1)_a$} :
      \begin{itemize}
        \item{$A_2$, etc\ldots}: If $r=0$ we have 
          $\ai\aix\aj=\aj=0\Rightarrow B=0$. So $r\not=0$, and we have :
          $$\aj\ai\aix=-q\ai\aix\aj$$ 
          $$\Rightarrow \aj=-q\aj$$ 
          Thus $q=-1$ or else $B=0$. Now, we can define $\phi :
          B\rightarrow C$, with
          $C=K\langle(x_i)_{i\in\II_\alpha}\rangle/\langle x_ix_j+x_jx_i-2\delta_{ij}\rangle$, by
          $\ai,\aix\mapsto x_i$, and we conclude by lemma 7.

        \item{$E$} : $r=0\Rightarrow
          \aj=\ai\aix\aj+\beta\aix\ai\aj=0-\beta\aix\aj\ai=0\Rightarrow B=0$.\newline
          We can then assume $r\not=0$. We have :
          $$\aj(\ai\aix+\beta\aix\ai)=\aj$$
          $$\Rightarrow -q(\ai\aix+\beta\aix\ai)\aj=\aj$$
          $$\Rightarrow (1+q)\aj=0$$
          Therefore $q=-1$.
           \begin{itemize}
             \item{$\beta\not=-1$} : We define $\phi : B\rightarrow
               K\langle(x_i)_{i\in\II_\alpha}\rangle/\langle x_ix_j+x_jx_i-{\displaystyle {2\over
               1+\beta}}\delta_{ij}\rangle$. We conclude as in case 
               $A_2$.
             \item{$\beta=-1$} : We have a pseudo-bosonic algebra $\hat A_\alpha$.
           \end{itemize}
        \item{$E\cup A_1\cup B_1$} : One can easily show that $r=s$ and 
          $1+\beta\not=0$. We can then map $B$ onto
          $C=K\langle x_i|i\in\II_\alpha\rangle/\langle
          x_ix_j+x_jx_i-2\delta_{ij}\rangle$ and use lemma 7.
      \end{itemize}
\medbreak
\underline{$(1,1)_c\cup(2,0)$} : The only case to study is $E$. In this case $B=0$ (see $(1,1)_c\cup(1,-1)_b\cup(2,0)$).
\medbreak
\underline{$(1,1)_{a\mbox{ or }b}\cup(2,0)$}:
       \begin{itemize}
       \item{$E$} : As we said at the beginning of this section, we have $\beta=1$. Therefore, we can define $\phi : B\rightarrow
         C_1:=K\langle a,a^+\rangle/\langle a^2,{a^+}^2,aa^++a^+a-1\rangle$, by setting $\ai\mapsto a$, $\aix\mapsto a^+$. We then use lemma 8.
       \item{$E\cup A_1\cup B_1$} : $\ai\aj=\ai(\ai\aix+\beta\ajx\aj)\aj=0$, so we are back to the case $(1,1)_c$.
       \end{itemize}

\underline{$(1,-1)_b\cup (2,0)$} :
        \begin{itemize}
          \item{$E$ or $E\cup A_1\cup B_1$} : $\beta=1$, and if $i\not=k\not=j$,
            $\ai(\ak\akx+\akx\ak)\aj=\lambda\ai\aj=0$, and we are back to a previous case.
        \end{itemize}

\underline{$(1,-1)_a\cup (2,0)$} :
        \begin{itemize}
          \item{$E$} : $\beta=1$. We can assume $r\not=0$, the case $s\not=0$ being symmetrical. $S=\{\ai\aix\r 1-\aix\ai,\ai^2\r 0,\aix^2\r 0,\ai\ajx\r q\ajx\ai\}$ is a confluent reduction system, which is adapted to $<_n$. Let $\JJ=(j_1,\ldots,j_r)$ and $\KK=(k_1,\ldots,k_s)$ be two tuples of indices. We write $\ajjx=a_{j_1}^+\ldots a_{j_r}^+$ and $\akk=a_{k_1}\ldots a_{k_s}$. By convention, $a_\emptyset=a_\emptyset^+=1$. With these notations, the basis of irreducible monomials is $T=\{\ajjx\akk\}$, with $\JJ$, $\KK$ running over all tuples such that $j_m\not=j_{m+1}$, $k_m\not=k_{m+1}$. The notation $k>\JJ$ shall mean that $k$ is greater than all indices appearing in $\JJ$. So let $k>\JJ,\KK$. We have :
              $$\matrix{
[\ajjx\akk,\ak] & = & \ajjx\akk\ak-\ak\ajjx\akk\hfill\cr
                & = & \ajjx\akk\ak-q^{|\JJ|}\ajjx\ak\akk\hfill\cr}$$
Then if lm$(N_i)$=$\ajjx\akk$, we see that for $k$ big enough : lm$([N_i,\ak])=\ajjx\ak\akk$, which is different from $-\ak$ and from 
$0$. Thus $(P_3)$ cannot hold.

            \item{$E\cup A_1\cup B_1$} : We have :
	$$\lambda\ai\aj=\ai(\az\azx+\beta\azx\az)\aj$$
	$$\Rightarrow\lambda\ai\aj=\ai\ai\aix\aj+\beta\ai\ajx\aj\aj$$
	$$\Rightarrow \ai\aj=0$$
        \end{itemize}

\underline{$(1,-1)_b$} : 
          \begin{itemize}
            \item{$E$,$E\cup A_1\cup B_1$} : With $k\not=i\not=l$, we have
              $\ak(\ai\aix+\beta\aix\ai)a^{}_l=0$.
           \end{itemize}

\underline{$(1,-1)_a$} : 
          \begin{itemize}
            \item{$A_2$} : If $r=0$ then $B=0$ (see
              $(1,-1)_a\cup(1,1)_a$). If $r\not=0$ then
              $S=\{\ai\aix\r 1,\ai\ajx\r q\ajx\ai\}$ is
              confluent and adapted to $<_n$. $T=\{\ajjx\akk\}$, where $\JJ$ and $\KK$ run over all tuples of indices, is the corresponding basis, and if $k>\JJ$, $\KK$ we have 
              : $[\ajjx\akk,\ak]=\ajjx\akk\ak-q^{|\JJ|}\ajjx\ak\akk$. Thus, we see that $(P_3)$ does not hold, as in case $(1,-1)_a\cup(2,0)\cup E$.
           \item{$A_2\cup B_1$, $A_2\cup B_2$} : $r\not=0$ by the above. We have, with $i\not=j$ :
             $$\aj\ajx\aj\ajx=1$$
             $$\Rightarrow\aj(\aix\ai)\ajx=q^2\aix\aj\ajx\ai=q^2\aix\ai=1$$
             therefore $q^2\ai\aix\ai=q^2\ai=\ai$, thus $q^2=1$ or else
             $B=0$.
             \begin{itemize}
               \item{If $q=1$} : We define $\phi : B\rightarrow
                 K[x,y]/\langle xy-1\rangle$, by $\phi(\ai)=x$, $\phi(\aix)=y$. We then use lemma 4.
               \item{If $q=-1$} : We define $\phi : B\rightarrow
                 K\langle(x_i)_{i\in\II_\alpha}\rangle/\langle x_ix_j+x_jx_i-2\delta_{ij}\rangle$,
                 and we use lemma 7.
             \end{itemize}
           \item{$E$} : We can assume $r\not=0$ (the case $s\not=0$ is symmetrical). $S=\{\ai\aix\r 1-\beta\aix\ai,\ai\ajx\r q\ajx\ai\}$ is confluent and adapted to $<_n$, and
                 $T=\{\ajjx\akk\}$. We find $[\ajjx\akk,\akx]=q^{|\KK|}\ajjx\akx\akk-\akx\ajjx\akk$, and we can do as above.
            \item{$E\cup A_1\cup B_1$} : Let us assume $r\not=0$, the case $s=0$ being symmetrical. We have :
$$(\ai\aix)\ai=\ai(\aix\ai)$$
$$\Rightarrow \ai-\beta\azx\az\ai=\ai\azx\az=q\azx\ai\az$$
$$\Rightarrow \ai\aix-\beta\azx\az\ai\aix=q\azx\ai\az\aix$$
$$\Rightarrow 1-\beta\azx\az-\beta\azx\az(1-\beta\azx\az)=q^2\azx(1-\beta\azx\az)\az$$
$$\Rightarrow 1+(-2\beta-q^2)\azx\az+\beta^2\azx(1-\beta\azx\az)\az=-\beta q^2\azx^2\az^2$$
\be
\Rightarrow 1+(\beta^2-2\beta-q^2)\azx\az+(-\beta^3+\beta q^2)\azx^2\az^2=0\label{for10}
\ee
Now we can define an homomorphism $\phi$ from the Weyl algebra $A_1:=K\langle a,a^+\rangle/\langle aa^+-a^+a-1\rangle$ to $B$ by $\phi(a)=\az$, and $\phi(a^+)=\azx$. Since $A_1$ is simple, either $\phi=0$ or $\phi$ is injective. But from (\ref{for10}), Ker$(\phi)\not=0$. Then $\phi=0$, therefore $B=0$.
          \end{itemize}

\underline{$(1,1)_c$} :
          \begin{itemize}
            \item{$E$, $E\cup A_1\cup B_1$} : With $j\not=i$, we have
              $\aj(\ai\aix+\beta\aix\ai)\aj\Rightarrow\aj^2=0$, and we find a case already studied.
          \end{itemize}

\underline{$(1,1)_b$} : 
          \begin{itemize}
            \item{$A_2$, etc\ldots} : We define $\phi : B\rightarrow
              K[x,y]/\langle xy-1\rangle$ by $\phi(\ai)=x$, $\phi(\aix)=y$, and we use lemma 5.
            \item{$E$ or $E\cup A_1\cup B_1$} : 
              \begin{itemize}
                \item{$\beta\not=-1$} : We define $\phi : B\rightarrow
                  K[x,y]/\langle xy-1/(1+\beta)\rangle$
                \item{$\beta=-1$} : We define $\psi : B\rightarrow
                  A_1$, by $\psi(\ai)=a$, $\psi(\aix)=a^+$, and we use lemma 8.
              \end{itemize}
          \end{itemize}

\underline{$(1,1)_a$} : 
           \begin{itemize}
             \item{$A_2$, etc\ldots} : We define $\phi :
               B\rightarrow
               K\langle(x_i)_{i\in\II_\alpha}\rangle/\langle x_ix_j+x_jx_i-2\delta_{ij}\rangle$, as in case $(1,1)_a\cup(1,-1)_a$.
             \item{$E$} :
               \begin{itemize}
               \item{$\beta\not=-1$} : We define $\phi$ as in case
                 $(1,1)_a\cup(1,-1)_a$.
               \item{$\beta=-1$} : We quotient out $B$ by the ideal generated by $\ai\aix-\az\azx$ and $\aix\ai-\azx\az$, we thus obtain the algebra $B'$ of case $E\cup A_1\cup B_1$. Now this algebra is non-zero and does not contain number operators (see below), so neither does $B$.
               \end{itemize}
             \item{$E\cup A_1\cup B_1$} : If $\beta\not=-1$ we do as above. If $\beta=1$ we see that the reduction system $S_0=\{\ai\aix\r 1+\azx\az,\ajx\aj\r\azx\az,\aj\ai\r-\ai\aj,\ajx\aix\r-\aix\ajx|i,j\in\II_\alpha,i<j\}$, adapted to $<_n$ is not confluent. Let us call $S$ the confluent reduction system that we get from $S_0$ by the non-commutative Buchberger algorithm (i.e. we inductively reduce every ambiguity. See \cite{Berg}, \cite{ufn} or \cite{bes1} for details on this algorithm). After three iterations (this can be calculated by hand, or preferably with a computer program, such as {\it bergman}, available at http://www.matematik.su.se/research/bergman/, or the one in \cite{bes1}, available at http://perso.wanadoo.fr/fabien.besnard/), one sees that $\forall i,j,k$, $\ai\aj\akx$ and $\ai\ajx\akx$ are reducible with respect to $S$ (all the details are in \cite{bes1}).

Now let $T$ be the basis of irreducible monomials corresponding to $S$, and let $x$ belong to $B^{(0,\ldots,0)}:=\{x\in B|$ $\forall i\in\II_\alpha, [N_i,x]=0\}$. According to what we have just said, $x$ is of the form : 
	\begin{itemize}
			\item{(1)} $a_{i_1}^+\ldots a_{i_k}^+a_{j_1}^{}a_{i_{k+1}}^+\ldots a_{j_r}^{}a_{i_{k+r}}^+a_{j_{r+1}}^{}\ldots a_{j_{r+k}}^{}$, or
			\item{(2)} $a_{i_1}^{}a_{j_1}^+\ldots a_{i_k}a_{j_k}^+$, or
			\item{(3)} $a_{i_1}^+a_{j_1}\ldots a_{j_k}^+a_{i_k}$
		    \end{itemize}
In each case, the tuple of indices $i$ and the tuple of indices $j$ are equal up to the order.

Since for all $i,j,k$, $\ai\aj\akx$ and $\ai\ajx\akx$ are reducible, one can see that the three cases reduce to : $x=\azx^k\az^k$. Thus :
$$B^{(0,\ldots,0)}=\{\azx^k\az^k|k\in\NN\}$$

As a consequence, we find that $\forall i,j$, $N_i\in B^{(0,\ldots,0)}$ is stable under the action of the transposition automorphism $\tau_{ij}^*$. Now :
$$\tau_{ij}^*[N_i,\aix]=[\tau_{ij}^*N_i,\tau_{ij}^*\aix]$$
$$\ssi \tau_{ij}^*\aix=[N_i,\ajx]$$
$$\ssi \ajx=0$$
Thus $(P_3)$ cannot hold. We must now prove that $B\not=0$.

We first note that the algebra $A$, generated by the $\xi_i$'s and $\xi_i^+$'s, satisfying $\xi_i^{}\xi_j^{}+\xi_j^{}\xi_i^{}=\xi_i^+\xi_j^++\xi_j^+\xi_i^+=0$, for $i\not=j$, and $\xi_i^{}\xi_i^+=\xi_i^+\xi_i^{}=1$, is non-zero (we can for instance quotient it by the ideal generated by the $\xi_i^{}-\xi_i^+$, thus obtaining a Clifford algebra which is clearly non-zero). Let us then consider $\bar B_1:=A\bigotimes A_1$, and set $\bi:=\xi_i^{}\otimes a$, $\bix:=\xi_i^+\otimes a^+$. We have, for $i\not=j$ :
$$\bi\bj+\bj\bi=\xi_i^{}\xi_j^{}\otimes a^2+\xi_j^{}\xi_i^{}\otimes a^2=(\xi_i^{}\xi_j^{}+\xi_j^{}\xi_i^{})\otimes a^2=0$$
$$\bix\bjx+\bjx\bix=\xi_i^+\xi_j^+\otimes {a^+}^2+\xi_j^+\xi_i^+\otimes {a^+}^2=(\xi_i^+\xi_j^++\xi_j^+\xi_i^+)\otimes {a^+}^2=0$$
$$\bix\bi-\bjx\bj=(\xi_i^+\xi_i^{})\otimes a^+a^{}-(\xi_j^+\xi_j^{})\otimes a^+a^{}=1\otimes a^+a^{}-1\otimes a^+a^{}=0$$
$$\bi\bix-\bj\bjx=(\xi_i^{}\xi_i^+)\otimes a^{}a^+-(\xi_j^{}\xi_j^+)\otimes a^{}a^+=1\otimes a^{}a^+-1\otimes a^{}a^+=0$$
and :
$$\bi\bix-\bix\bi=(\xi_i^{}\xi_i^+)\otimes a^{}a^+-(\xi_i^+\xi_i^{})\otimes a^+a^{}=1\otimes(a^{}a^+-a^+a^{})=1\otimes 1=1$$
We thus see that $B$ can be non-trivialy mapped to $\bar B_1$ by $\psi(\ai):=\bi$ and $\psi(\aix)=\bix$, which proves that $B\not=0$.
\smallbreak
           \end{itemize}

\underline{$(2,0)$} :
          \begin{itemize}
            \item{$E$ or $E\cup A_1\cup B_1$} : We already know that $\beta=1$. We can thus define $\phi$ as in case $(1,1)_b\cup(2,0)$.
          \end{itemize}

\underline{$\emptyset$} : In each sub-case, we can define an homomorphism $\phi$ as in one of the cases above.
          \begin{itemize}
            \item{$A_2$, etc\ldots} : $B\rightarrow K[x,y]/\langle xy-1\rangle$
            \item{$E$, $E\cup A_1\cup B_1$}
              \begin{itemize}
                \item{$\beta\not=-1$} : $B\rightarrow
                  K[x,y]/\langle xy-1/(1+\beta)\rangle$
                \item{$\beta=-1$} : $B\rightarrow A_1$
              \end{itemize}
          \end{itemize}
\bigbreak
So we see that the only possibilities were (pseudo)-fermions and (pseudo)-bosons, that we have respectively found in cases $(1,1)_{a\mbox{ or }b}\cup(1,-1)_a\cup(2,0)\cup E$, and $(1,1)_{a\mbox{ or }b}\cup(1,-1)_a\cup E$.

\section{Topological Number Operator Algebras}
\subsection{Definitions}
In order to emcompass the case of $q$-bosons, we must relax the conditions we impose on number operator algebras so as to let the number operators belong to some completion of the algebra. We are led to the following definition :
\begin{definition}\label{topnoa}
Let $B$ be a non-trivial $K$-algebra. Let $X_A=\{\ai|i\in\II_\alpha\}$ and $X_{A^+}=\{\aix|i\in\II_\alpha\}$ be 2 sets of distinct elements of $B$. Let $V_n:=BX_A^n$ be the left ideal of $B$ generated by $X_A^n$ and let $\tilde B:=\limproj_{n\in\NN^*}B/V_n$. If the following conditions hold :
\item{($H_1$)} : ${\displaystyle \bigcap_{n\in\NN^*}}V_n=\{0\}$
\item{($H_2$)} : $\forall b\in X_A^+, \forall n\in\NN, \exists N\in\NN$ such that $V_Nb\subset V_n$

then the canonical morphism $B\r\tilde B$ is an embedding of algebras. If, in addition to this, we have :
\item{(i)} $B$ is generated by $X_A\cup X_{A^+}$ as an algebra.
\item{(ii)} One uniquely defines an anti-involution $J$ on $B$ by setting $J(\ai)=\aix$.
\item{(iii)} For all $i\in\II_\alpha$, there exists $\tilde N_i\in\tilde B$ such that for all $j\in\II_\alpha$ : 
\be
[\tilde N_i,\aj]=-\delta_{ij}\aj\label{eq5}
\ee
\be
[\tilde N_i,\ajx]=\delta_{ij}\ajx\label{eq6}
\ee
then $(B,X_A,X_{A^+},(N_i)_{i\in\II_\alpha})$ is called a topological number operator algebra, and $\tilde B$ is the completion of $B$ for the topology generated by the neighbourhoods $V_n$ of the origin.
\end{definition}

In order for this definition to make sense, we must prove a few things. First, since the $V_n$'s are left ideals of $B$ satisfying $V_m\subset V_n$ whenever $m\geq n$ they form a projective system of left ideals and $\tilde B$ is a left $B$-module. We also know that they form a basis of neighbourhoods of zero of a topology for which the sum and the left mulitplication are continuous.
The property $(H_1)$ shows that the canonical map is into and that the topology is separated. We still have to show that the multiplication is a continuous mapping from $B\times B$ to $B$. For this, it is easy to see that we only need to show the continuity of right multiplication. This is assured by the property $(H_2)$ for multiplication by elements of $X_{A^+}$. Since it is obviously true for elements of $X_A$, we can show that it is true for any monomial, and then for any element of $B$.

With this definition we can expect the elements of $\tilde B$ to be expressed as normal ordered series, that is to say with the creations to the left.

In the next lemmas, $B$ is a topological n.o.a.
\begin{lemma}
Let $S_0=\{0\}$ and $\forall k\in\NN^*$, $S_k\supset S_{k-1}$ a subspace of $B$ supplementary to $V_k$. Then $\tilde B$ is isomorphic to the set of power series of the type 
$$S=\sum_{k\in\NN}u_k\mbox{ with }u_k\in S_{k+1}\cap V_k$$
endowed with the obvious laws. 
\end{lemma}
\dem\newline
Let us define the natural projections $\pi_k : B\rightarrow B/V_k$, and $\pi_{k,j} : B/V_k\rightarrow B/V_j$, for $j\leq k$. Every $x\in\tilde B$ is given by a sequence $(x_k)_{k\in\NN^*}$ such that $x_k\in B/V_k$ and $\forall j\leq k$, $\pi_{k,j}(x_k)=x_j$. Let $s_k$ be the linear section of $\pi_k$ associated with $S_k$. We set $u_k=s_{k+1}(x_{k+1})-s_k(x_k)$, and $S(x)=\sum_{k\in\NN}u_k$. Conversely if $S=\sum_k u_k$ we set $x_k=\pi_k(\sum_{j=0}^{k-1}u_j)$.

In it trivial to verify that we have defined two linear maps, inverse to each other. Indeed, the lemma is just a restatement of the definition of the projective limit, with in addition the condition $(H_2)$ assuring that the product of two series is well defined.\hfill QED.
\smallbreak
\underline{Remark} : If it happens that $V_n=V_{n+1}$ for some $n$, then $V_m=V_n$ for every $m\geq n$. We have $S_m=S_n$ and the series are just finite sums. Thus, in this case $\tilde B$ is embedded in $B$. From $(H_1)$ we finally get $B=\tilde B$, and $V_n=\{0\}$.
\smallbreak

Let us see now a particular case.

\begin{lemma}\label{ray}
Set $T=\{a_{i_1}^+\ldots a_{i_k}^+a_{j_1}\ldots a_{j_l}|k,l\geq 0\}$ and $T_l=\{a_{i_1}^+\ldots a_{i_k}^+a_{j_1}\ldots a_{j_l}|k\geq 0\}$. Suppose that $T$ generates $B$ as a $K$-space. Then there exist $T_0'\subset T_0$, $T_1'\subset T_1$, \ldots such that for all $k$, $T_0\coprod\ldots\coprod T_k'$ is a basis of $B/V_{k+1}$.

Moreover, every $x\in\tilde B$ can be written in a unique way :
$$x=\sum_{l=0}^\infty\sum_{t\in T_l'}\lambda_{t,l}t$$
with the condition that $\forall l$, the set $\{t\in T_l'|\lambda_{t,l}\not=0\}$ is finite.
\end{lemma}
\dem\newline
Since $V_k=BX_A^k$ and $T$ is a generating family for $B$, we have $V_k=$Span$\{T_j|j\geq k\}$. We choose for $T_0'\subset T_0$ a basis of $B/V_1$. Then we choose $T_1'$ so that $T_0'\coprod T_1'$ is a basis of $B/V_2$, and so on. Then we set $S_k=$Span$\{T_j'|j<k\}$ and apply the previous lemma.\hfill QED.
\smallbreak
\underline{Remark} : It is not assumed that $T'=\bigcup_{n\in\NN}T_n'$ is a basis of $B$. In fact this assumption implies $(H_1)$ and seems to be strictly stronger.
\smallbreak

Of course a topological n.o.a. $B$ is $\ZZ^{\II_\alpha}$-graded, and so will be its completion $\tilde B$. For every $n\in\ZZ^{\II_\alpha}$, we write $\tilde B^n=\{x\in\tilde B|\forall i\in\II_\alpha$ $[N_i,x]=n(i)x\}$.

\begin{lemma}\label{lemme21}
Let $n\in\ZZ^{\II_\alpha}$. Then $\forall x\in\tilde B^n$, $\exists (x_k)\in B^{\NN^*}$ such that for all $k$ $x_k\in B^n$ and $x=\lim_k x_k$.
\end{lemma} 
\dem\newline
Let $B':=\bigoplus_{p\in\ZZ^{\II_\alpha},p\not=n}B^p$. $\forall k\in\NN^*$, $\exists z_k\in B$ such that $x-z_k\in V_k$. Now $z_k=x_k+y_k$ with $x_k\in B^n$ and $y_k\in B'$. Since $V_k$ is stable under ad$(N_i)$ for all $i$, we have $x-x_k\in V_k$ and $y_k\in V_k$.\hfill QED.
\smallbreak

To give another motivation for the definition \ref{topnoa}, let us introduce a new kind of algebra, that would seem more natural in a physicist's point of view : 

\begin{definition}\label{def4}
Let $B$ be a non-trivial $K$-algebra satisfying $(i)$ and $(ii)$ of definition \ref{topnoa}. Let us call $F$ the left module $F=B/BV_1$ and let us define $\rho : B\r$End$(F)$ such that $\rho(x)(u)=x.u$. If the following properties are satisfied :
\item{(a)} Ker$(\rho)=\{0\}$.
\item{(b)} $F$ is generated by the monoid $X_{A^+}^*$ as a $K$-space.
\item{(c)} $\forall i\in\II_\alpha$, $\exists N_i\in\End(F)$ s.t. 
\be
[N_i,\rho(\aj)]=-\delta_{ij}\rho(\aj)\label{com6}
\ee
\be
[N_i,\rho(\ajx)]=\delta_{ij}\rho(\ajx)\label{com7}
\ee
\item{(d)} $N_i(|1\rangle)=0$, where $|1\rangle=1 [V_1]$.
\smallbreak
then $(B,X_A,X_{A^+},(N_i)_{i\in\II_\alpha})$ is called a Fock algebra and we call $\rho$ the Fock representation of $B$.
\end{definition}
Thanks to (\ref{com6}) and (\ref{com7}), a Fock algebra is graded in exactly the same way as a n.o.a., and we can define $B^p$ and the $i$-numbers for $i$-homogenous elements.

In the next lemmas, $B$ is a Fock algebra.
\begin{lemma}\label{lem13}
Let $i\in\II_\alpha$ and $x\in B$ such that $x$ is $i$-homogenous and $n_i(x)\leq 0$. Then $\exists\lambda\in K$ such that $x=\lambda$ $[V_1]$. Furthermore, if $n_i(x)<0$, $\lambda=0$.
\end{lemma}
\dem\newline
By $(b)$ of definition (\ref{def4}), there exists a linear combination of elements of $X_{A^+}^*$, let us say $\lambda+\xi$, where $\lambda$ is the scalar part, such that $x-\lambda-\xi\in V_1$. The result is then clear using the fact that $V_1$ is graded.\hfill QED.

\begin{lemma}\label{lem14}
$\forall i,j\in\II_\alpha$, $\forall k\in\NN^*$, $\exists$ $y_{ij}^1,\ldots,y_{ij}^{k-1}$ such that $y_{ij}^n\in X_{A^+}^nX_A^n$, $y_{ij}^n$ has the same numbers as $\ai\ajx$ and $\ai\ajx=\sum_{n=0}^{k-1}y_{ij}^n$ $[V_k]$.
\end{lemma}
\dem\newline
By the previous lemma the result holds for $k=1$. If it holds for $k$, then $\ai\ajx=\sum_{n=0}^{k-1}y_{ij}^n+v_k$, with $v_k\in V_k$. Write $v_k=\sum w_tz_t$ $[V_{k+1}]$ with $w_t\in B$ and $z_t\in X_A^k$. Then each $w_t$ can be decomposed as a linear combination of some elements in $X_{A^+}^*$ modulo $V_1$. Thus we can write $v_k=\sum w_t'z_t$ $[V_{k+1}]$ with $w_t'\in X_{A^+}^*$. Now the numbers of $w_t'z_t$ are the same as those of $\ai\ajx$. Thus $z_t$ must contain $\ai$ and $w_t'$ must contain $\ajx$, at least once. If we remove one copy of these two generators in $z_t$ and $w_t'$ respectively,  we find that the numbers of the remaining two monomials must cancel. Since one consists of generators only, and the other of destructions only, they must be of the same length. So $w_t'\in X_{A^+}^k$.\hfill QED.

\begin{proposition}
If $B$ is a Fock algebra, then $B$ fulfills $(H_1)$ and $(H_2)$.
\end{proposition}
\dem\newline
Let $x\in\bigcap_{n\in\NN^*}V_n$ and $m$ be a monomial. Take $n$ larger than the length of $m$, and any $n$-uple $\II$. Then $n_i(\aii m)<0$, so $\aii m=0$ $[V_1]$ by lemma \ref{lem13}. Thus $V_nm=0 [V_1]$ and $\rho(x)(m)=0$, consequently $x\in$Ker$(\rho)=\{0\}$, and $(H_1)$ is fulfilled.

Let us show that $V_k\aix\subset V_{k-1}$ for all $i$, which clearly entails $(H_2)$. This is true for $k=1$. If it is true for $k$, then $\forall p\leq k-1$, $V_kX_{A^+}^p\subset V_{k-p}$. Let $x$ be a monomial in $V_{k+1}$. Write $x=x'\aj$. We have $\aj\aix=\sum_{p=0}^{k-1}\lambda_{k,p}v_{k,p}w_{k,p}+n_k$, with $n_k\in V_k$, $v_{k,p}\in X_{A^+}^p$ and $w_{k,p}\in X_A^p$, by lemma \ref{lem14}. Thus $x'v_{k,p}\in V_{k-p}$ and $x\in V_k$. So by induction, we find the claimed result.\hfill QED.
\smallbreak
\underline{Remark} : Let $B$ be a topological n.o.a. and a Fock algebra, and put the discrete topology on $F$. Then the topology of $\tilde B$ is stronger than the topology of pointwise convergence on End$(F)$. In all the cases we will investigate, these topologies are in fact the same.

\subsection{The classification theorem} 

\begin{theorem}\label{theo3}
If $B=L_\alpha/I$ is a topological n.o.a. of type $\alpha$ that is symmetric and quadratically presented, then either $I$ is one of the ideals enumerated in theorem \ref{theo2} or there exist $h\in R\setminus\{0\}$, $q\in R\setminus\{-1,1\}$ such that $I$ is generated by
	\begin{itemize}
	\item{(d)} $\{\ai\aj-\aj\ai,\aix\ajx-\ajx\aix,\ai\ajx-\ajx\ai,\ai\aix-q\aix\ai-1|i\not=j\}$
	\item{(d')} $\{\ai\aj+\aj\ai,\aix\ajx+\ajx\aix,\ai\ajx+\ajx\ai,\ai\aix-q\aix\ai-1|i\not=j\}$
	\end{itemize}
In the case $(d)$, the algebra is called a $q$-boson algebra, and denoted by $A_\alpha^q$. In the case $(d')$ it is called a pseudo-$q$-boson algebra, and denoted by $\hat A_\alpha^q$. In both cases we have ${\displaystyle N_i=\sum_{k=0}^\infty{(1-q)^k\over 1-q^k}\aix^k\ai^k+\lambda_i}$, $\lambda_i\in K$, the algebras $\tilde B$ are central and the number operators are unique up to an additive constant. 
\end{theorem}

\subsection{Sketch of proof of theorem \ref{theo3}}

Let $B$ be an algebra fulfilling the hypotheses of theorem \ref{theo3}. This algebra will also fulfill the hypotheses of theorem \ref{theo2}, except $(P_3)$, which is replaced by $(\tilde P_3)$ which is the same as $(P_3)$ except that the $N_i$ are allowed to belong to $\tilde B$ instead of $B$. One can verify that proposition \ref{prop4} still holds in this context, since it only depends on the gradation of $B$ by the $i$-numbers. Thus all we have to do is to re-examine the cases of section 5.3 which have been eliminated by the hypothesis $(P_3)$. We will have then to verify whether they fulfill $(H_1)$, $(H_2)$, and $(\tilde P_3)$.
\smallbreak

First of all, let us note that in the cases containing $B_2$ we have $\aix^n\ai^n=1$, thus $(H_1)$ cannot be true.

Let us look at the remaining cases.
\smallbreak
\underline{$(1,1)_b\cup(1,-1)_a\cup(2,0)$} :
\begin{itemize}
	\item{$E$, $\beta=1$} : We have the pseudo-fermionic algebra $B=C_\alpha$. The hypotheses of lemma \ref{ray} are fulfilled, $(H_1)$ and $(H_2)$ are easily seen to be satisfied. The elements of $\tilde B$ can be written
$$\sum_{l=0}^\infty\sum_{i_1<\ldots<i_k\atop j_1<\ldots<j_l}\lambda_{i_1,\ldots,i_k}^{j_1,\ldots,j_l}a_{i_1}^+\ldots a_{i_k}^+a_{j_1}^{}\ldots a_{j_l}$$
the second sum being finite for each $l$.\newline
If $x\in Z(\tilde B)$, $x=\lim_k x_k$, with $x_k\in B^0$, by lemma \ref{lemme21}, so that we must have $\forall i\in\II_\alpha$, $\lim[x_k,\ai]=0$.

Thus $\forall n$, $\exists p$, $k\geq p\Rightarrow [x_k,\ai]\in V_n$. Now $[a_{i_1}^+\ldots a_{i_k}^+a_{j_1}\ldots a_{j_k},\ai]$ equals $a_{i_1}^+\ldots a_{i_{s-1}}^+a_{i_{s+1}}^+\ldots a_{i_k}^+a_{j_1}\ldots a_{j_k}$, if $i=i_s$, and $0$ if $i\notin\{i_1,\ldots,i_k\}$.

We thus have $x_k\in K+V_n+V_{\II\setminus\{i\}}$, where $V_{\II\setminus\{i\}}$ is the left ideal generated by $\{\ajx|j\not=i\}$. So we see that $x\in K+V_{n}+V_{\II\setminus\{i\}}$, for all $n$ and for all $i$. This shows that $x\in K$.
\end{itemize}
\medbreak
\underline{$(1,1)_a\cup(1,-1)_a\cup(2,0)$} : This case is similar to the previous one, except that for $E$, $\beta=1$, we get $B=\hat C_\II$.
\medbreak
\underline{$(1,1)_b\cup(1,-1)_a$} :
	\begin{itemize}
		\item{$A_2$} : See $E$, with $\beta=0$.
		\item{$A_2\cup B_1$} : We must have $r+s=0$. As in the finite case one shows that $\aix\ai=1$. Thus $(H_1)$ cannot be fulfilled.
		\item{$E$} : $r+s=0$. We are in the bosonic and $q$-bosonic ($q=-\beta$) cases. We can use lemma \ref{ray} with $T=\{\aiix\ajj|i_1\leq\ldots\leq i_k,\ j_1\leq\ldots\leq j_l\}$. $(H_2)$ is clearly true. One shows that $\tilde B$ is central exactly as we did in case $(1,1)_b\cup(1,-1)_a\cup(2,0)$. For $q\not=\pm 1$, one sees that the number operators are given by
$$N_i=\sum_{k=0}^\infty{(1-q)^k\over 1-q^k}\aix^k\ai^k+\lambda_i.1$$
(this had been shown in \cite{cs}).
		\item{$E\cup A_1\cup B_1$} : $r+s=0$. As we have already seen, we have $Z(B)=B^0$. Then if $(H_1)$, $(H_2)$ and $(\tilde P_3)$ were true, we would have $Z(\tilde B)=\tilde B^0$, which entails that $(\tilde P_3)$ cannot be true, a contradiction.
	\end{itemize}
\medbreak
\underline{$(1,1)_a\cup(1,-1)_a$}: This case is similar to the previous one. We find the pseudo-$q$-bosons in case $E$.
\medbreak
\underline{$(1,1)_c\cup(2,0)$} : Since $\forall k\geq 2$, $V_k=\{0\}$, we have $\tilde B\simeq B$.
\medbreak
\underline{$(1,1)_{a\ or\ b}\cup(2,0)$}
	\begin{itemize}
		\item{$E$}: $\beta=1$. We define $\phi:$ $B\rightarrow C_1$ as in 2.4. Suppose that there exist number operators $\tilde N_i\in\tilde B$. Then let $u_n\in B$ be such that $N_i=\lim_nu_n$. Take $j\not=i$. We must have $\lim_n([u_n,\ai]+\ai)=0$ and $\lim_n[u_n,\aj]=0$. Thus $\exists n$ such that $[u_n,\ai]+\ai\in V_2$ and $[u_n,\aj]\in V_2$. But $\phi(V_2)=0$. Thus $[\phi(u_n),a]+a=0$ and $[\phi(u_n),a]=0$, which is absurd.
		\item{$E\cup A_1\cup B_1$} : same as above.
	\end{itemize}
\medbreak
\underline{$(1,-1)_a\cup(2,0)$}
	\begin{itemize}
		\item{$E$}: $\beta=1$. 
			\begin{itemize}
			\item{$r\not=0$}: The reduction system $S=\{\ai\aix\r 1-\aix\ai,\ai^2\r 0,\aix^2\r 0,$ $\ai\ajx\r q\ajx\ai|i\not=j\}$ is confluent. The basis of irreducible monomials is $T=\{\aiix\ajj|i_r\not=i_{r+1}$ and $j_s\not=j_{s+1}\}$. Let us define the degree of a monomial in $B$ to be the degree of its normal form. Then $d^\circ(\ajjx)=k$, and $\ai\ajjx$ is a linear combination of monomials of degree $\geq k$. More precisely, they are of degree $k+1$ whenever $i\notin\JJ$.\newline
Suppose there exists a number operator $N_i$. Then $N_i=\sum_k N_i^k$, with $d^\circ(N_i^k)=2k$. Now let $j$ be such that $j\not=i$ and $j$ does not appear in any monomial of the support of $N_i^1$ or $N_i^2$ (there is a finite number of such monomials). Then $[N_i,\aj]=0\Rightarrow[N_i^1,\aj]=0$. Now $N_i^1=\sum_k\lambda_k\akx\ak$ (finite sum). Thus $[N_i^1,\aj]=\sum_k\lambda_k(\akx\ak\aj-q\akx\aj\ak)=0$, so $\lambda_k=0$ and $N_i^1=0$.\newline
Now only $[N_i^1,\ai]$ can provide terms of degree $1$ in $[N_i,\ai]$. Consequently, $N_i$ cannot exists.
			\item{$r=0$}: Let us multiply $a_{i_1}^+$ to the right by $a_{i_2}^{}a_{i_2}^++a_{i_2}^+a_{i_2}^{}$, with $i_1\not=i_2$. We get : $a_{i_1}^+=a_{i_1}^+a_{i_2}^+a_{i_2}$. In the same way we have $a_{i_2}^+=a_{i_2}^+a_{i_3}^+a_{i_3}$, with $i_3\not=i_2$, thus $a_{i_1}^+=a_{i_1}^+a_{i_2}^+a_{i_3}^+a_{i_3}a_{i_2}$. By induction we see that $a_{i_1}^+\in\bigcap_{k\in\NN^*}V_k$, so $(H_1)$ cannot be fulfilled.
			\end{itemize}
	\end{itemize}
\medbreak
\underline{$(1,-1)_a$} :
	\begin{itemize}
		\item{$A_2$}, $r=0$ : We do as in the case $(1,-1)_a\cup(2,0)\cup E$.
		\item{$A_2\cup B_1$}, $r\not=0$ : we find $\aix\ai=1$.
		\item{$E$} :
			\begin{itemize}
			\item{$r\not=0$} : We have a confluent reduction system and we can use the same argument as in case $(1,-1)_a\cup(2,0)$.
			\item{$r=0$} : Since $\ajx\aix\ai=\ajx$, we can do as in case $(1,-1)_a\cup(2,0)$.
			\end{itemize}
	\end{itemize}
\medbreak
\underline{$(1,1)_b$} :
	\begin{itemize}
		\item{$A_2$} : See $E$, $\beta=0$.
		\item{$A_2\cup B_1$} : The following relations hold in $B$ :
$$\matrix{
\azx\az\ajx=\ajx,&\forall j\not=0\cr
\ajx\ak\aix=\aix\ak\ajx,&\forall j>i,\ k\not=i,j\cr
\ai\aj\aix=\aj,&\forall i<j\cr
\ai\aj\akx=\ai\akx\aj,&\forall i\leq j,k\not=i,j\cr
\ak\aix\ajx=\aix\ak\ajx,&\forall i\leq j,k\not=i,j\cr
\aj\aix\ajx=\aix,&\forall i<j\cr
\aj\akx\ai=\ai\akx\aj,&\forall j>i,\ k\not=i,j\cr
\aj\azx\az=\aj,&\forall j\not=0\cr}$$
To see it, one should use the non-commutative Buchberger algorithm on the initial reduction system. This can be done with the computer programs already cited. See also \cite{bes1} for a detailed account of the calculations.
So, as in 2.4, $(1,1)_a\cup E\cup A_1\cup B_1$, we can prove that $B^0=$Span$\{\azx^k\az^k|k\in\NN\}$. Now take $i$ and $j$, two distinct indices, and set $N_i=\lim_n x_n$. We have : $\forall n\in\NN$, $\exists k$ such that $[x_k,\ai]+\ai\in V_n$ and $[x_k,\aj]\in V_n$. Now if $\tau_{i,j}^*$ is the automorphism induced by the $(i,j)$-transposition, we have : $\tau_{i,j}^*x_k=x_k$, and $\tau_{i,j}^*V_n=V_n$, so that $\tau_{i,j}^*[x_k,\aj]=[x_k,\ai]\in V_n$, thus $\ai\in V_n$, and this is true for all $n$. So if $(H_1)$ is satisfied, $(\tilde P_3)$ is not.
		\item{$E$ or $E\cup A_1\cup B_1$}:
			\begin{itemize}
			\item{$\beta=-1$} : We can send $B$ onto $A_1$ by $\psi$, as in the finite case. Suppose that $N_i$ is a number operator, and take $j\not=i$. If $u_n$ is a sequence in $B$ that converges towards $N_i$, we must have for $n$ large enough : $[u_n,\ai]+\ai\in V_1$ and $[u_n,\aj]\in V_1$, so that $[\phi(u_n),a]+a\in a^+A_1$ and $[\phi(u_n),a]\in a^+A_1$. Now $(-a+a^+A_1)\cap a^+A_1=\emptyset$, a contradiction.
			\item{$\beta\not=-1$} : We can use the same method if we send $B$ onto the $q$-bosonic algebra ($q=-\beta$) $A_1^q$. Since $(-a+a^+A_1^q)\cap A_1^q=\emptyset$, we arrive at the same conclusion.
			\end{itemize}
	\end{itemize}
\medbreak
\underline{$(1,1)_a$} : 
	\begin{itemize}
		\item{$A_2$} : Let us prove the following relations by induction :
$$\ai a_{i_1}^{}\ldots a_{i_j}^{}\aix=(-1)^ja_{i_1}^{}\ldots a_{i_j}^{}$$
$$\ak a_{i_1}^+\ldots a_{i_j}^+\akx=(-1)^ja_{i_1}^+\ldots a_{i_j}^+$$
for all $i<i_1<\ldots<i_j<k$. They are true for $j=0$, so let us suppose they are true for $j$. It suffices to multiply on the left by $a_{i_{j+1}}$ in the first relation, and on the right by $a_{i_{j+1}}^+$ in the second one, to see that they are still true for $j+1$. It is then easy to see that the reduction system formed by $\ai a_{i_1}^{}\ldots a_{i_j}^{}\aix\r (-1)^ja_{i_1}^{}\ldots a_{i_j}^{}$, $\ak a_{i_1}^+\ldots a_{i_j}^+\akx\r(-1)^ja_{i_1}^+\ldots a_{i_j}^+$, and $\aj\ai\r -\ai\aj$, $\ajx\aix\r -\aix\ajx$ for all $j>i$, is confluent. So $(H_2)$ is not satisfied, indeed : $\forall i_1\leq\ldots\leq i_k$ and $i\notin\{i_1,\ldots,i_k\}$, $\ai a_{i_1}^+\ldots a_{i_k}^+$ is irreducible and $V_n=$Span$\{\ajjx x|$ $|\JJ|=n$ and $\ajjx x$ irreducible$\}$.
		\item{$A_2\cup B_1$} : We can use the same method as in $(1,1)_b$ (see \cite{bes1} for the computer calculations).
		\item{$E$} : Let us call $p$ the quotient map $p:$ $B\rightarrow B':=B/\langle\{\ai\aix-\az\azx,\aix\ai-\azx\az|i\in\II\}\rangle$. Suppose $B$ satisfies $(\tilde P_3)$ and set $N_i=\lim_n x_n$, $i\not=0$, with $x_n\in B^0$. Then $p(x_n)\in {B'}^0$ but we know that ${B'}^0$ is generated by the $\azx^k\az^k$'s. Thus $p(x_n)$ is invariant by any transposition $\tau_{ij}^*$. For all $k$ there exists an $n$ such that $[x_n,\ai]+\ai\in V_k$, and $[x_n,\aj]\in V_k$. Thus $[p(x_n),p(\ai)]+p(\ai)\in p(V_k)$ and $[p(x_n),p(\aj)]\in p(V_k)$. Using $\tau_{ij}^*$ on the second equation, and substracting from the first, we find $p(\ai)\in p(V_k)$. 
Then, using $\psi : B'\rightarrow\bar B_\beta$, defined in the paragraph 5.3 :
$$a\otimes\xi_i\in(a^+\otimes 1)\bar B_\beta$$
but this is false, as we can see by using the basis $\{{a^+}^ra^s\otimes\xi_t\}$, where $\{\xi_t\}$ is a basis of $A$ containing $\xi_i$.
		\item{$E\cup A_1\cup B_1$} : We know that $B^0=$Span$\{\azx^k\az^k\}$, thus we can do as in case $(1,1)_b\cup A_2\cup B_1$.\newline

	\end{itemize}

\section{Concluding Remarks}
In this article, we have tried to explore the algebraic constraints that a free field theory must abide by. Of course this approach have raised as many questions as it has answered. Imposing quadratic relations (and confluence in the finite case) seems to be just as restrictive as needed in order to state a classification theorem. In this way we have recovered all the known cases, plus a new one if the number of degrees of freedom is finite. The virtue of this method is also to put on an equal footing bosons, fermions, pseudo-bosons and pseudo-fermions, which shows that $\epsilon$-symmetry (see \cite{bes2}) appears in a natural way.

There are at least two directions towards which we can try to go further : incorporating infinite sums in the defining relations and allowing cubic relations in order to recover para-statistics. These subjects are under investigations but what we have done so far indicates that other algebraic hypotheses must be imposed to keep the problem feasible.

\end{document}